\documentclass[twocolumn]{aastex61}

\usepackage{multirow}

\shorttitle{Mergers Might Not be the Only Source}
\shortauthors{C\^ot\'e et al.}

\begin{document}


\title{Neutron Star Mergers Might not be the Only Source of\\r-Process Elements in the Milky Way}

\correspondingauthor{Benoit C\^ot\'e}
\email{benoit.cote@csfk.mta.hu}

\author[0000-0002-9986-8816]{Benoit C\^ot\'e}
\affil{Konkoly Observatory, Research Centre for Astronomy and Earth Sciences, Hungarian Academy of Sciences, Konkoly Thege Miklos ut 15-17, H-1121 Budapest, Hungary}
\affiliation{National Superconducting Cyclotron Laboratory, Michigan State University, East Lansing, MI 48824, USA}
\affiliation{Joint Institute for Nuclear Astrophysics - Center for the Evolution of the Elements, USA}
\affiliation{NuGrid Collaboration, \url{http://nugridstars.org}}
 
\author{Marius Eichler}
\affiliation{Institut f\"ur Kernphysik, Technische Universit\"at Darmstadt, Schlossgartenstr. 2, Darmstadt 64289, Germany}

\author{Almudena Arcones}
\affiliation{Institut f\"ur Kernphysik, Technische Universit\"at Darmstadt, Schlossgartenstr. 2, Darmstadt 64289, Germany}
\affiliation{GSI Helmholtzzentrum f\"ur Schwerionenforschung GmbH, Planckstr. 1, Darmstadt 64291, Germany}

\author{Camilla J. Hansen}
\affiliation{Max Planck Institute for Astronomy, Koenigstuhl 17, 69117 Heidelberg, Germany}

\author{Paolo Simonetti}
\affiliation{Astronomy Department, University of Trieste, Via Tiepolo 11, I-34127 Trieste, Italy}

\author{Anna Frebel}
\affiliation{Department of Physics and Kavli Institute for Astrophysics and Space Research, Massachusetts Institute of Technology, Cambridge, MA
02139, USA}
\affiliation{Joint Institute for Nuclear Astrophysics - Center for the Evolution of the Elements, USA}

\author{Chris L. Fryer} 
\affiliation{Center for Theoretical Astrophysics, LANL, Los Alamos, NM 87545, USA}
\affiliation{Department of Astronomy, The University of Arizona, Tucson, AZ 85721, USA}
\affiliation{Department of Physics and Astronomy, The University of New Mexico, Albuquerque, NM 87131, USA}
\affiliation{The George Washington University, Washington, DC 20052, USA}
\affiliation{Joint Institute for Nuclear Astrophysics - Center for the Evolution of the Elements, USA}
\affiliation{NuGrid Collaboration, \url{http://nugridstars.org}}

\author{Marco Pignatari}
\affiliation{E.A. Milne Centre for Astrophysics, Department of Physics \& Mathematics, University of Hull, HU6 7RX, United Kingdom.}
\affiliation{NuGrid Collaboration, \url{http://nugridstars.org}}
\affiliation{Joint Institute for Nuclear Astrophysics - Center for the Evolution of the Elements, USA}
\affil{Konkoly Observatory, Research Centre for Astronomy and Earth Sciences, Hungarian Academy of Sciences, Konkoly Thege Miklos ut 15-17, H-1121 Budapest, Hungary}

\author{Moritz Reichert}
\affiliation{Institut f\"ur Kernphysik, Technische Universit\"at Darmstadt, Schlossgartenstr. 2, Darmstadt 64289, Germany}

\author{Krzysztof Belczynski}
\affiliation{Nicolaus Copernicus Astronomical Center, Polish Academy of Sciences, ul. Bartycka 18, 00-716 Warsaw, Poland}

\author{Francesca Matteucci}
\affiliation{Astronomy Department, University of Trieste, Via Tiepolo 11, I-34127 Trieste, Italy}
\affiliation{I.N.A.F. Osservatorio Astronomico di Trieste, via G.B. Tiepolo 11, I-34131, Trieste, Italy}
\affiliation{I.N.F.N. Sezione di Trieste, via Valerio 2, I-34134 Trieste, Italy}

\begin{abstract}

Probing the origin of r-process elements in the universe represents a multi-disciplinary challenge.  We review the observational evidence that probe the properties of r-process sites, and address them using galactic chemical evolution simulations, binary population synthesis models, and nucleosynthesis calculations. Our motivation is to define which astrophysical sites have significantly contributed to the total mass of r-process elements present in our Galaxy.  We found discrepancies with the neutron star (NS-NS) merger scenario.  Assuming they are the only site, the decreasing trend of [Eu/Fe] at [Fe/H]\,$>-1$ in the disk of the Milky Way cannot be reproduced while accounting for the delay-time distribution (DTD) of coalescence times ($\propto~t^{-1}$) derived from short gamma-ray bursts and population synthesis models. Steeper DTD functions ($\propto~t^{-1.5}$) or power laws combined with a strong burst of mergers before the onset of Type~Ia supernovae can reproduce the [Eu/Fe] trend, but this scenario is inconsistent with the similar fraction of short gamma-ray bursts and Type~Ia supernovae occurring in early-type galaxies, and reduces the probability of detecting GW170817 in an early-type galaxy.  One solution is to assume an extra production site of Eu that would be active in the early universe, but would fade away with increasing metallicity. If this is correct, this extra site could be responsible for roughly 50\,\% of the Eu production in the early universe, before the onset of Type~Ia supernovae. Rare classes of supernovae could be this additional r-process source, but hydrodynamic simulations still need to ensure the conditions for a robust r-process pattern.

\end{abstract}

\keywords{Galaxy: abundances -- Star: abundances -- Physical Data and Processes: nuclear reactions, nucleosynthesis, abundances -- Binaries: close}




\section{Introduction}
\label{sec:intro}
Understanding the origin of rapid neutron-capture process (r-process) elements in the universe requires a multi-scale framework including nuclear astrophysics, stellar spectroscopy, gravitational waves, short gamma-ray bursts (GRBs), and galaxy formation theories (e.g., \citealt{2005ARA&A..43..531B,2007PhR...450...97A,2014ARA&A..52...43B,2015ARA&A..53..631F,2016ARNPS..66...23F,2017ApJ...848L..13A,2017ARNPS..67..253T,2018arXiv180608955F,2018arXiv180504637H,2019arXiv190101410C}). To best interpret stellar abundances of r-process elements\footnote{By r-process elements we mean neutron-capture elements that are, to the best of our knowledge, mostly produced by the r-process.} derived from spectroscopy, nucleosynthesis calculations must be convolved with galaxy evolution simulations. To include r-process sites in such simulations, the general properties of those sites must be known, which in the case of neutron star (NS-NS) mergers can be constrained by gravitational wave and short GRBs detections.

The goal of our study is to build a coherent inter-disciplinary picture to identify which astrophysical site(s) have significantly contributed to the amount of r-process elements present in the Milky Way. We review the different observational evidence that probe the properties of r-process sites, and investigate them from the perspective of galactic chemical evolution (GCE) simulations, binary population synthesis models, and theoretical nucleosynthesis. In particular, we focus on the tensions that emerge when NS-NS mergers are assumed to be the only r-process site (see Table~\ref{tab_summary}). We refer to \cite{2018ApJ...869...50D} and Sk{\'u}lad{\'o}ttir et al.~(in preparation) for a discussion of r-process sites in dwarf galaxies.

Many studies addressed the challenge of explaining with NS-NS mergers the presence of Eu, a lanthanide element, in the atmosphere of the most metal-poor stars ([Fe/H]\,$\lesssim$\,$-2$) of the Galactic halo (e.g., \citealt{2004A&A...416..997A,2014MNRAS.438.2177M,2015A&A...577A.139C,2015ApJ...804L..35I,2015ApJ...807..115S,2015MNRAS.452.1970W,2018arXiv181202779S,2019MNRAS.483.5123H}, see also Section~\ref{sec:scenario_5}). In this paper, we instead focus on the decreasing trend of [Eu/Fe]\footnote{[A/B]~$\equiv$~log$_{10}(n_\mathrm{A}/n_\mathrm{B})-$\,log$_{10}(n_\mathrm{A}/n_\mathrm{B})_\odot$, where $n_\mathrm{A}$ and $n_\mathrm{B}$ represent the number density in the stellar atmosphere of elements A and B, respectively. The second term on the right-hand side represents the solar composition.} at [Fe/H]\,$>$\,$-1$ in the Galactic disk. Currently, this trend cannot be well reproduced by invoking a proper delay-time distribution (DTD) for NS-NS mergers (\citealt{2017ApJ...836..230C,2018arXiv180101141H,2019arXiv190102732S}).  This distribution can be seen as the probability of a merger event to occur at a given delay time following the formation of the binary system.

In this paper, we explore different DTD functions for NS-NS mergers in order to recover the evolution of [Eu/Fe] in the Galactic disk using chemical evolution models. We then discuss whether those new DTD functions are consistent with other constraints outside the world of chemical evolution, such as the host galaxies of short GRBs, gravitational wave detections, and binary population synthesis predictions. We also present a list of alternative r-process enrichment scenarios in the Milky Way with multiple r-process sites involving rare classes of supernova, and discuss their implications with respect to a diverse range of observational and theoretical constraints. Overall, this work aims to isolate and highlight the current inter-disciplinary tensions that need to be addressed in future studies in order to identify the dominant r-process site(s) in the Milky Way.

The outline of this paper is as follows.  We review the observational evidence related to the production of r-process elements in Section~\ref{sec:obs}.  In Sections~\ref{sect_NSNS_GCE}, \ref{sec:pop_synth}, \ref{sec:nucleo}, we address the properties of r-process sites from the point of view of galactic chemical evolution simulations, binary population synthesis models, and theoretical nucleosynthesis calculations, respectively.  In Section~\ref{sec:challenges}, we highlight the inter-disciplinary tensions associated with existing r-process enrichment scenarios involving compact binary mergers and magneto-rotational supernovae, and propose alternative scenarios to be confirmed or disproved by future work.  Our discussion and conclusions are in Sections~\ref{sec:disc} and \ref{sec:conclusions}, respectively.

\section{Observational Evidence}
\label{sec:obs}
Here we review the observational evidence that probe the properties of the r-process sites. We interpret these observables with numerical simulations in Sections~\ref{sect_NSNS_GCE} and \ref{sec:nucleo}, and attempt to build a consistent multi-disciplinary picture in Section~\ref{sec:challenges}.

\subsection{Neutron-Capture Elements in Metal-Poor Stars}
\label{sec:obs_met_poor}
Metal-poor stars in the Galactic halo and in dwarf galaxies can be unique tracers of r-process nucleosynthesis in the early universe (e.g., 
\citealt{2008ARA&A..46..241S,2018arXiv180608955F,2018arXiv180504637H}).  Throughout this paper, we will refer to the three peaks in the r-process abundance distribution, the first peak at Se-Kr, the second peak at Te-Xe, and the third peak at Os-Pt-Au. Heavier than the third peak are the actinide elements such as Th and U. We note  that the solar r-process residual pattern is obtained by subtracting the contribution of the slow neutron-capture process (s-process) from the total solar abundances (e.g., \citealt{kaeppeler:89,arlandini:99,Burris2000,Simmerer2004,2008ARA&A..46..241S,Bisterzo2014}).

\subsubsection{r-Process Enhanced Stars}
\label{sec:obs_robustness}
For stars with strong enhancements in neutron-capture elements, abundances of more than 35 elements between Sr and U can be derived (\citealt{2012ApJS..203...27R,2012ApJ...750...76R,Mello2013,2014A&A...569A..43M,2018ApJ...856..138J}), typically with accuracies better than $\pm$\,0.2\,dex, owing to large telescopes, high resolution spectrographs, and improved knowledge of stellar parameters and atomic physics (e.g., \citealt{Sneden2003,Hansen2015,2017ApJ...847..142E}). Some metal-poor stars are moderately or strongly enhanced in r-process elements and are classified based on their [Eu/Fe] and [Ba/Eu] ratios: r-I stars when $\mbox{[Eu/Fe]}>0.3$ and $\mbox{[Ba/Eu]}<0$, r-II stars when $\mbox{[Eu/Fe]}>1.0$ and $\mbox{[Ba/Eu]}<0$. We note that Ba is actually past the second r-process peak. Eu is located between the second and third r-process peak.

\cite{Sneden2003} first showed that the abundances of CS~22892-052, an r-II star, was similar within $0.2-0.3$\,dex to the solar r-process pattern between the second and third peak. Since then, many studies found additional r-process enhanced stars that also show this pattern with similar precisions (\citealt{2009ApJ...698.1963R,2011rrls.conf..223C,2012ApJS..203...27R,2014A&A...569A..43M,2016ApJ...830...93J,2017ApJ...844...18P,2018arXiv180511925H,2018ApJ...854L..20S}). This pattern has also been found in dwarf galaxy stars, such as in Reticulum~II (\citealt{2016ApJ...830...93J,2018ApJ...856..138J}). This universal abundance pattern, between the second and third r-process peak, is referred to as the main r-process. Deriving abundances for elements \textit{in} the second r-process peak requires space telescopes with spectrographs operating in the ultraviolet. This has been done for a handful of stars only, and their abundances are consistent with the main portion of the scaled solar r-process pattern (\citealt{2012ApJ...750...76R,2012ApJ...747L...8R}).

While the main r-process observationally appears very robust, r-process enhanced stars do show large variations up to $\sim$\,1\,dex among first-peak neutron-capture elements such as Sr, Y, and Zr (e.g., \citealt{Hansen2014,2014MNRAS.445.2946R,2016ApJ...830...93J,2018ApJ...858...92H,2018ApJ...856..138J,Spite2018}). Although the origin of those variations is still unclear, it suggests there is more than one astrophysical site that can produce the first-peak elements (e.g., \citealt{Hansen2014,2015A&A...577A.139C,Spite2018}). At low metallicity, these elements could be made by different nucleosynthesis processes (see, e.g., \citealt{2007ApJ...671.1685M,2010ApJ...724..975R}) such as neutrino-driven winds and the $\nu$p-process in CC~SNe \citep{{froehlich:06,farouqi:09,roberts:10,Arcones2011,2013JPhG...40a3201A}}, the s-process in fast-rotating massive stars \citep{{pignatari:08, frischknecht:16}}, or the intermediate neutron-capture process (i-process, \citealt{roederer:16}).

Besides variations in the first peak, there is also a range of abundances found in the actinide elements with atomic number $Z\geq80$ (e.g., \citealt{2002A&A...387..560H,roederer:09}). Stars with an high Th/Eu ratio relative to the solar r-process pattern have been termed actinide-boost stars. \cite{2018ApJ...856..138J} recently reported the discovery of an actinide-deficient star in Reticulum\,II. This has broadened the range of Th abundances relative to the solar pattern to $\sim$\,0.6\,dex (see also \citealt{schatz2002,2009ApJ...698.1963R,2014A&A...569A..43M,2018ApJ...855...83H,2018arXiv180511925H}). We note that U abundances scale with Th values, but only very few U measurements are available. It is currently difficult to determine whether the variations in actinide abundances is a result of different r-process sites, or the trace of different ejecta with different physical conditions within the same site (see discussion in Section~\ref{sec:disc_actinides}).

\subsubsection{Limited r-Process Stars}
\label{sec:obs_limited_r}
In contrast to the r-process enhanced stars, the majority of very metal-poor stars with $\mbox{[Fe/H]}<-2.0$ do not show any significant enhancements in neutron-capture elements. Although their abundances or upper limits among halo stars are often extremely low, they still show large star-to-star scatter. Some of these stars display no super-solar enhancement in neutron-capture elements but nevertheless display a systematic depletion of Ba and heavier elements relative to the abundances of lighter neutron-capture elements such as Sr. Stars with this signature has recently been termed limited r-process stars \citep{2018arXiv180608955F}. The metal-poor star HD~122563 (\citealt{2006ApJ...643.1180H}) is a famous example. However, as for the large scatter in the abundances of first-peak elements in r-process enhanced metal-poor stars (see Section~\ref{sec:obs_robustness}), the nucleosynthetic origin(s) of limited r-process stars is still unclear.

\subsubsection{Scatter of [Eu/Fe] in Galactic Halo Stars}
\label{sec:obs_scatter}
\begin{figure}
\center
\includegraphics[width=3.35in]{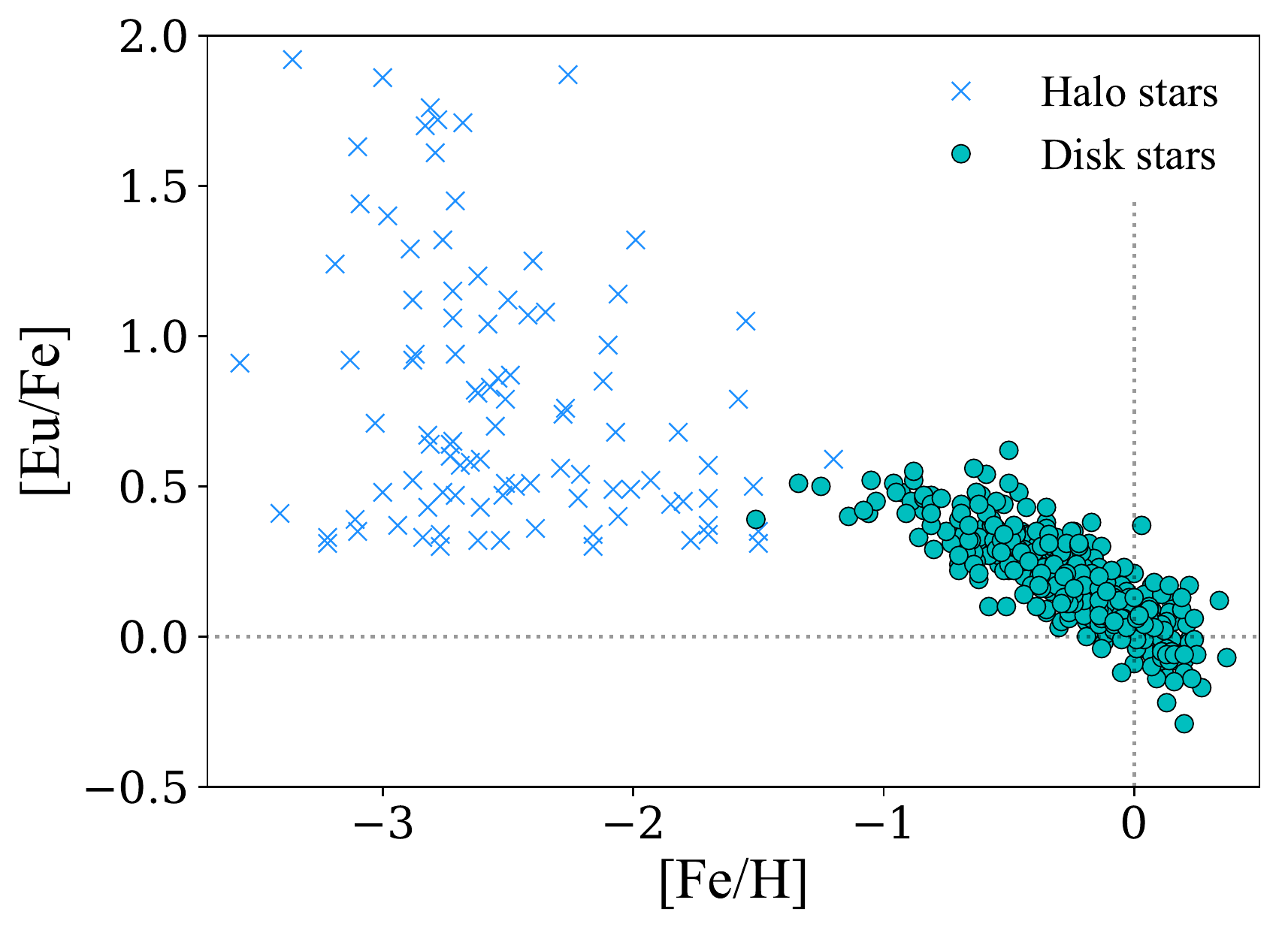}
\caption{Evolution of europium abundances ([Eu/Fe]) as a function of iron abundances ([Fe/H]) observed in the Milky Way. The blue crosses represent a compilation of data extracted with the JINABase \citep{2017arXiv171104410A}, containing \cite{Burris2000}, \cite{Sneden2003}, \cite{2004ApJ...603..708C}, \cite{2005A&A...439..129B}, \cite{2007ApJ...660L.117F}, \cite{2013AJ....145...13A}, \cite{2014ApJ...797...69R}, and \cite{2015ApJ...807..173H}. We excluded upper limits and only selected stars with $[$Eu/Fe$]>0.3$ and $[$Ba/Eu$]<0$. The latter criteria aims to remove metal-poor s, i, and r+s stars.  To this compilation, we added the 12 r-II stars found in \cite{2018ApJ...858...92H}. The cyan circles are disk star data taken from \cite{2016A&A...586A..49B}. The dotted black lines mark the solar values (\citealt{2009ARA&A..47..481A}). \label{fig:Eu_data}}
\end{figure}

The large scatter seen in the [Eu/Fe] ratio of r-process enhanced metal-poor stars in the Milky Way halo (blue crosses in Figure~\ref{fig:Eu_data}), compared to the smaller scatter seen in [$\alpha$/Fe]\footnote{The $\alpha$ elements are mostly produced by massive stars and include elements such as O, Mg, Si, Ca, and Ti.}, indicates that the production of Eu in the early universe must have been rare and prolific compared to the production of $\alpha$ elements by standard CC~SNe (e.g., \citealt{2015A&A...577A.139C,2015ApJ...814...41H,2015MNRAS.452.1970W,2018MNRAS.tmp..585N}). The recent analysis of \cite{2018ApJ...860...89M} suggests a minimum ejection of $\sim$\,10$^{-3}$\,M$_\odot$ of r-process material per event to explain the most Eu-enhanced stars. The rarity of r-process events is also necessary to explain the low frequency of r-process enhanced ultra-faint dwarf galaxies such as Reticulum~II (\citealt{2016Natur.531..610J,2016AJ....151...82R}). A comparison between the abundance of $^{244}$Pu in the Earth's ocean crust and the value derived for the early solar system also supports the idea that the r-process elements should be produced in rare events that eject a large amount of mass \citep{wallner2015,2015NatPh..11.1042H,LUGARO20181}.

\subsection{Evolution of [Eu/Fe] in the Galactic Disk} 
\label{sec:obs_trends}
At higher metallicity ([Fe/H]\,$>$\,$-1.5$) in the Galactic disk, there is no star with a clean r-process signature because of the production of neutron-capture elements by the s-process (e.g., \citealt{1998ApJ...497..388G}). To probe the evolution of r-process elements in disk stars, we use Eu. According to the solar r-process pattern, defined by subtracting the s-process pattern from the total solar abundance pattern, Eu was produced at $\sim$\,97\,\% by the r-process (e.g., \citealt{Burris2000}), at the time the solar system formed 4.6\,Gyr ago (\citealt{2017GeCoA.201..345C}). We note, however, that there are uncertainties in s-process yields (e.g., \citealt{Bisterzo2014,2015ApJ...801...53C}), and that the i-process could potentially alter the exact composition of the r-process residuals by contributing to the production of neutron-capture elements (see e.g., \citealt{Dardelet2014,Hampel2016,2016ApJ...821...37R,2018ApJ...854..105C,2018JPhG...45e5203D}). We discuss the possible production of Eu by the i-process in Section~\ref{sec:disc_i_proc}. Throughout this paper, we assume that Eu is a reliable r-process tracer.

An important feature of the chemical evolution of Eu in the Milky Way is the decreasing trend of [Eu/Fe] in the disk, as recorded in stars with [Fe/H]~$>$~$-1$ (cyan dots in Figure~\ref{fig:Eu_data}).  This data behaves similar to the chemical evolution of [$\alpha$/Fe] (see e.g., \citealt{2014A&A...562A..71B,2018MNRAS.tmp.1218B,2018MNRAS.474.2580S}).  Because $\alpha$ elements are mostly produced in massive stars (e.g., \citealt{2002RvMP...74.1015W,2013ARA&A..51..457N}), we can assume that the overall production rate of r-process elements should also follow the lifetime of massive stars.  In particular, because this decreasing trend originates from the extra production of Fe by Type~Ia supernovae (SNe~Ia) that mixes with the material ejected by CC~SNe (e.g., \citealt{1986A&A...154..279M,2009A&A...501..531M,2014MNRAS.444.3845F}), the production of r-process elements should occur on timescales shorter than SNe~Ia, as for $\alpha$ elements (see Section~\ref{sect_NSNS_GCE} for more details). We note that the AMBRE Project recently measured abundances for Gd and Dy, two other tracer elements of the r-process, in the Galactic disk (\citealt{2018A&A...619A.143G}). They showed that [Gd/Fe] and [Dy/Fe] also have decreasing trends at [Fe/H]~$>$~$-1$.

\subsection{Short Gamma Ray Bursts and Type~Ia Supernovae}
\label{sec:obs_GRB}
While long-duration GRBs are associated with the explosion of massive stars, the progenitors of short-duration GRBs are believed to be NS-NS or black hole~-~neutron star (BH-NS) mergers (see \citealt{2014ARA&A..52...43B} for a review).  The detection of the short GRB that followed the NS-NS merger event GW170817 measured by LIGO/Virgo (\citealt{2017ApJ...848L..13A}) reinforced even more this idea.  We discuss the broad implication of GW170817 in Section~\ref{sec:obs_grav}.  For now, we focus on the properties of short GRBs to probe the coalescence timescales of compact binary systems. We also compare results from short GRB observations with those of SNe~Ia observations in order to investigate whether those two types of events occur on similar timescales. This is relevant for GCE simulations (see Section~\ref{sect_NSNS_GCE}).

\subsubsection{Host Galaxies}
\label{sec:obs_GRB_host}
In the review of \cite{2014ARA&A..52...43B}, $\sim$\,30\,\% of the 26 short GRBs with classified host galaxies are found in early-type galaxies (see also \citealt{2013ApJ...769...56F,2015JHEAp...7...73D}). A similar fraction of $\sim$\,$1/3$ has been derived by \cite{2017ApJ...848L..23F} with the 36 short GRBs detected between 2004 and 2017. Early-type (giant elliptical and S0) galaxies typically show a lack or low level of recent star formation compared to late-type (spiral) galaxies (e.g., \citealt{2016A&A...590A..44G}) and are dominated by old stellar populations (e.g., \citealt{2018arXiv180407769V}).  This implies that a certain fraction of NS-NS mergers, if they are the source of short GRBs, must have long coalescence timescales.  Otherwise, they would be exclusively found in star-forming late-type galaxies, as for core-collapse supernovae (CC~SNe) (e.g., \citealt{2011MNRAS.412.1473L}) and long GRBs (e.g., \citealt{2014ARA&A..52...43B}).

A fraction of SNe~Ia is also observed in early-type galaxies (e.g., \citealt{2005A&A...433..807M,2011MNRAS.412.1473L}).  Using the $\sim$\,370 classified SNe~Ia found in the Lick Observatory Supernova Search (LOSS), between 15\,\% and 35\,\% of SNe~Ia occur in early-type galaxies (Figure~5 in \citealt{2011MNRAS.412.1419L}, see also \citealt{2017ApJ...837..120G,2017ApJ...837..121G}), depending on whether we exclude or include S0-type galaxies in the early-type category.  Similar percentages of 10\,\% and 32\,\% are obtained using the 103 SNe~Ia with classified host galaxies from the Carnegie Supernova Project (\citealt{2017AJ....154..211K}). Using $\sim$\,450 classified SNe~Ia found in the Sloan Digital Sky Survey-II (SDSS-II) Supernova Survey\footnote{\url{https://data.sdss.org/sas/dr10/boss/papers/supernova/}} (\citealt{2014arXiv1401.3317S}), between 12\,\% and 40\,\% of SNe~Ia are found in galaxies with stellar masses above $0.5-1.0\times10^{11}$\,M$_\odot$ (see also \citealt{2017ApJ...850..135U}), which is roughly the mass range above which early-type galaxies become the dominant type in galaxy populations (e.g., \citealt{2014MNRAS.444.1647K,2016MNRAS.457.1308M,2016MNRAS.459...44T}). Using the same sample, we find that $\sim$\,25\,\% of SNe~Ia occur in passive galaxies that do not form stars anymore.

Compared to the $\sim$\,30\,\% derived for short GRBs with classified host galaxies, this suggests that short GRBs and SNe~Ia occur on average on similar timescales. 

\subsubsection{Delay-Time Distributions}
\label{sec:obs_GRB_DTD}
The DTD function of an astronomical event represents the probability of that event to occur at a given time $t$ following the formation of its progenitor objects.  Different approaches for deriving observationally the DTD function of SNe~Ia are described in \cite{2014ARA&A..52..107M}. Regardless of the different methodologies, most studies point toward a DTD in the form of $t^{-1}$ (\citealt{2008PASJ...60.1327T,2010ApJ...722.1879M,2012MNRAS.426.3282M,2011MNRAS.417..916G,2014ApJ...783...28G,2012AJ....144...59P}), which is consistent with the predictions of population synthesis models (e.g., \citealt{2009ApJ...699.2026R}).

Similar techniques can be applied to short GRBs.  According to \cite{2017ApJ...848L..23F}, and references therein, the DTD function of short GRBs could also be in the form of $t^{-1}$.  This is also in agreement with population synthesis studies (e.g., \citealt{2012ApJ...759...52D,2018MNRAS.474.2937C}).  As mentioned in Section~\ref{sec:obs_GRB_host}, the statistics of short GRB detections is still low and the derived DTD function could change in the near future (see also discussion in \citealt{2017ApJ...848L..23F}).  We note that \cite{2015JHEAp...7...73D} argued for a steeper DTD function in the form of $t^{-1.5}$ for short GRBs.  However, this provides tension with respect to the similar fraction of short GRBs and SNe~Ia occurring in different types of galaxies (see Section~\ref{sec:obs_GRB_host}), unless the DTD function of SNe~Ia turns out to also be in the form of $t^{-1.5}$, as suggested by the recent analysis of \cite{2017ApJ...834...15H}.

According to the 13 confirmed NS-NS binary systems observed in the Milky Way compiled by \cite{2017ApJ...846..170T}, six of them have an estimated coalescence time. This excludes systems that will merge in more than 50\,Gyr. The additional NS-NS binary PSR J1946+2052 discovered by \cite{2018ApJ...854L..22S} has an estimated coalescence time of 46\,Myr. When put in an increasing order, these seven NS-NS binaries should merge in 46, 86, 217, 301, 480, 1660, and 2730\,Myr.  Although a larger sample is needed to derive a reliable DTD function, the fact that two of the seven systems have coalescence timescales larger than 1\,Gyr is consistent with the idea that the number of NS-NS mergers, if they are responsible for short GRBs, should be distributed following a long-lasting DTD function.

\subsection{The Gravitational Wave GW170817}
\label{sec:obs_grav}
The detection of GW170817 by LIGO/Virgo (\citealt{2017ApJ...848L..13A}) and its associated electromagnetic emissions is to date the best direct evidence that NS-NS mergers can synthesize r-process elements.  In the next subsections, we review the implication of this event on the contribution of NS-NS mergers on the origin of r-process elements in the Milky Way.

\subsubsection{The Host Galaxy NGC\,4993}
\label{sec:obs_grav_host}
This first NS-NS merger detection occurred in NGC\,4993, an early-type galaxy (\citealt{GW170817-EM,2017Sci...358.1556C}).  Analysis of the properties of this galaxy revealed a current stellar mass of about $0.3-1.4\times10^{11}$\,M$_\odot$ and a star formation rate (if any) below $\sim$\,10$^{-2}$\,M$_\odot$\,yr$^{-1}$ (\citealt{2017ApJ...848L..22B,2017ApJ...849L..16I,2017ApJ...848L..28L,2017ApJ...848L..30P}), which is significantly lower compared to what is observed in late-type galaxies having similar stellar masses (see Figure~4 in \citealt{2016A&A...590A..44G}).  According to \cite{2017ApJ...848L..28L}, an upper limit of only $\sim$\,1\,\% of the stellar mass could originate from young stars.  Given the properties of the host galaxy of GW170817, \cite{2017ApJ...848L..30P} argued that the delay between the formation of the progenitor binary system and the NS-NS merger event have likely been greater than $\sim$\,3\,Gyr.  By convolving the star formation history of NGC\,4993 with a DTD function in the form of $t^{-1}$, \cite{2017ApJ...848L..22B} calculated the probability of the merger timescale to be between 6.8 and 13.6\,Gyr, with 90\,\% confidence.

According to population synthesis models, the DTD function of NS-NS mergers does extend up to $\sim$\,10\,Gyr, but most of the merger events are likely to occur within the first Gyr following the formation of the progenitor stars (e.g., \citealt{2018MNRAS.474.2937C}). Therefore, according to \cite{2017arXiv171200632B}, detecting the first NS-NS merger event in an early-type galaxy was unlikely, although not impossible. The low probability of this event is also discussed in \cite{2017ApJ...848L..28L}.  With only one NS-NS gravitational wave detection, conclusions are limited, but the host galaxy of GW170817 does inform us that it is at least possible for NS-NS mergers to have Gyr-long coalescence timescales. We refer to \cite{2017ApJ...849L..34P} for a discussion on a possible formation mechanism for the binary neutron star system that led to GW170817 in NGC\,4993.

\subsubsection{The Kilonova AT 2017gfo (SSS17a)}
\label{sec:obs_grav_kilo}
The ultraviolet, optical, and infrared emission from GW170817 suggest a significant production of r-process elements (e.g, \citealt{chornock17,cowperthwaite2017,drout2017,pian2017a,tanaka17b,2017ApJ...851L..21V}).  In particular, its late-time infrared emission suggests the production of lanthanide elements (e.g., \citealt{2013ApJ...775...18B,2016ARNPS..66...23F,tanvir2017,2018MNRAS.478.3298W}, but see \citealt{2018A&A...615A.132R,2018ApJ...868...65W,2018arXiv180810459W}).  Multi-dimensional simulations are based on two-component models (e.g., \citealt{tanvir2017}). The first component is the neutron-rich dynamical ejecta that produces heavy r-process elements up to the third peak, at the time the compact objects collide. The second one is a wind component primarily composed of first-peak r-process elements, which is launched at later time once the compact objects have merged. We refer to Appendix~\ref{sec_nucleo_nsm} for more details on those two components and on their nucleosynthesis.

All models, however, used approximate opacities since detailed databases of opacities for all heavy elements are not yet available.  State-of-the-art studies used a few opacities as surrogates for the entire rare-Earth lanthanide element distribution \citep{2013ApJ...775...18B,2018MNRAS.478.3298W}, but many others used constant opacities as a function of temperature, density, and wavelength.  In some calculations, the inferred ejected mass of heavy r-process elements is less than $10^{-5}$\,M$_\odot$ \citep{arcavi2017}.  But most models predicted $\sim$\,$0.001-0.01$\,M$_\odot$ of  dynamical ejecta, and $\sim$\,$0.01-0.03$\,M$_\odot$ of wind ejecta (see Table~1 in \citealt{2018ApJ...855...99C}).

The ejected mass of the dynamical and wind components both depend on the total mass and the mass ratio of the merging neutron stars.  Typically, systems with extreme mass ratios eject more dynamical mass~\citep[][and references therein]{korobkin2012,2017PhRvD..96l4005B} and produce larger disk masses \citep{2013ApJ...762L..18G}.  If the wind ejecta represent a constant fraction of the disk mass, we would expect the wind ejecta to also increase with more extreme mass ratios, although the exact amount of ejected mass depends on the simulation~\citep[see][and references therein]{2018ApJ...858...52S}.  The current constraint on the neutron star mass ratio for GW170817 ranges from 0.4 to 1.0 (\citealt{2017PhRvL.119p1101A}).

Although the wind ejecta are thought to mostly be composed of first-peak r-process elements, they may also include some heavier elements. This could alter the ratio of the abundances between the second and third r-process peaks, an undesired effect given the robustness of the r-process between the second and third peaks, as observed in r-process enhanced metal-poor stars (see Section~\ref{sec:obs_robustness}).  

\subsubsection{The Merger Rate Density}
\label{sec:obs_grav_rate}
The local NS-NS merger rate density of 1540$^{+3200}_{-1220}$ Gpc$^{-3}$\,yr$^{-1}$ provided by LIGO/Virgo (\citealt{2017PhRvL.119p1101A}) represents a significant step forward in constraining the role of NS-NS mergers on the production of r-process elements in the universe. Several analytical calculations, based on estimates for the total mass of r-process elements currently present in the Milky Way, showed that the rate could be high enough to explain all the r-process mass with NS-NS mergers only (\citealt{abbott17c,chornock17,cowperthwaite2017,2017arXiv171005442G,kasen2017,rosswog2017,tanaka17b,2017arXiv171005805W,2018arXiv180101141H}, see also \citealt{1999A&A...341..499R}).  A similar conclusion was derived using a compilation of galactic chemical evolution simulations (\citealt{2018ApJ...855...99C}).

Unfortunately, the uncertainties in the merger rate and in the total mass ejected per NS-NS merger (see Section~\ref{sec:obs_grav_kilo}) remain very large and it is not possible at the moment to come to a firm conclusion regarding the actual contribution of NS-NS mergers (see Figure~3 in \citealt{2018ApJ...855...99C}). Therefore, although NS-NS mergers are likely to be an important site of r-process nucleosynthesis, their existence does not rule out possible contributions by other r-process sites, such as rare classes of CC~SNe and BH-NS mergers.

\section{Chemical Evolution Simulations}
\label{sect_NSNS_GCE}
GCE simulations can bridge nuclear astrophysics efforts and stellar abundances derived from spectroscopy.  In addition to the yields and properties of different enrichment sites, GCE simulations also take into account galaxy evolution processes such as star formation and interactions between galaxies and their surrounding medium (e.g., \citealt{2003PASA...20..401G,2008EAS....32..311P,2013ARA&A..51..457N,2014SAAS...37..145M,2015ARA&A..53...51S}).  In this section, we describe how GCE simulations can be used to test the impact of NS-NS mergers on the evolution of r-process elements in the Milky Way. The goal is not to use GCE simulations to define what is the dominant r-process site. The goal is rather to use those simulations to provide a constraint that must be combined with other constraints coming from the other fields of research covered in this paper.

There are two approaches commonly used to include NS-NS mergers in GCE simulations.  The first one is to assume that all NS-NS mergers occur after a constant delay time following the formation of the NS-NS binaries (e.g., \citealt{2004A&A...416..997A,2014MNRAS.438.2177M}).  The second approach is to assume that NS-NS mergers in a given stellar population are distributed in time according to a long-lasting DTD function typically in the form of $t^{-1}$ (e.g., \citealt{2015ApJ...807..115S,2015MNRAS.447..140V}). In the later approach, NS-NS mergers can have a wide range of coalescence times, from a few tens of Myr to several Gyr (e.g., \citealt{2015IJMPD..2430012R}). We explore alternative forms of DTD functions in Section~\ref{sec:scenario_3}.

\begin{figure}
\center
\includegraphics[width=3.35in]{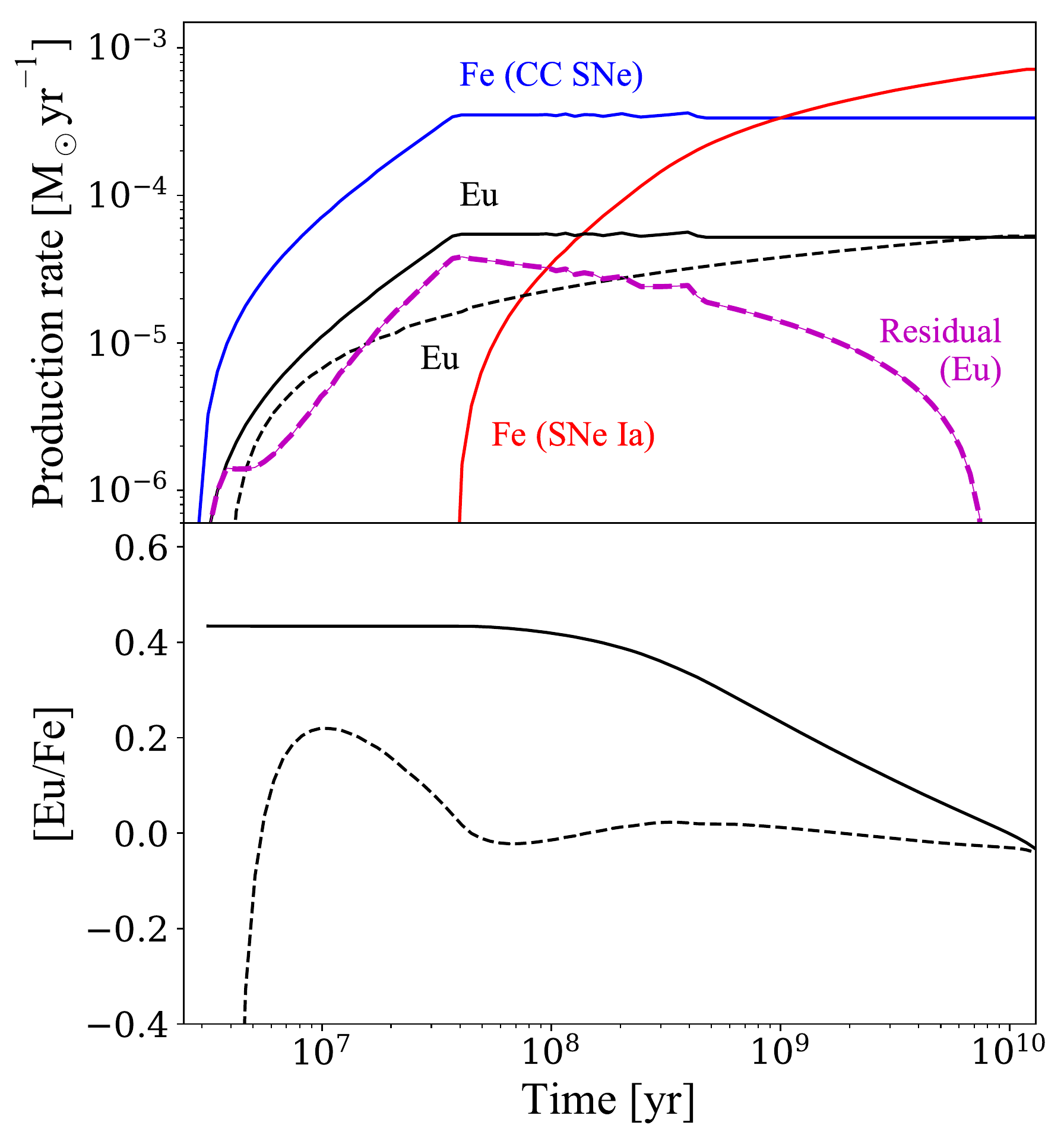}
\caption{Depiction of how the decreasing trend of [Eu/Fe] can be generated with time. {\bf Top panel:} Production rates of Fe in CC~SNe (blue line) and SNe~Ia (red line) as a function of time in a galaxy with a constant star formation rate of 1\,M$_\odot$\,yr$^{-1}$.  The black lines show the hypothetical production rate of Eu (scaled up by a factor of $4\times10^5$ for visualization purposes) when assuming that Eu is ejected following the lifetime of massive stars (solid black line) or a delay-time distribution function in the form of $t^{-1}$ (dashed black line), similar to one adopted for SNe~Ia. The pink residual dashed line is a possible extra source of Eu in the early universe (discussed in Section~\ref{sec:scenario_5}), obtained by subtracting the black dashed line from the solid black line. {\bf Bottom panel:} Evolution of [Eu/Fe] using the two different assumptions used in the top panel for the production timescale of Eu (same line styles). Results have been calculated with a simplified version of \texttt{OMEGA} (see Section~\ref{sect_knee}), assuming $3.35\times10^{-4}$\,M$_\odot$ of ejected Fe per units of stellar mass formed for CC~SNe (see \citealt{2017ApJ...836..230C}), 0.7\,M$_\odot$ of ejected Fe per SN~Ia, and $10^{-5}$\,M$_\odot$ of ejected Eu per r-process event. \label{fig:knee}
}
\end{figure}

\subsection{Reproducing the Decreasing Trend of [Eu/Fe]}
\label{sect_knee}
The bottom panel of Figure~\ref{fig:knee} shows the predicted evolution of [Eu/Fe] as a function of galactic age assuming a constant delay time (black solid line) and a long-lasting DTD function ($\propto t^{-1}$, black dashed line) for NS-NS mergers, assuming they are the only source of Eu. These predictions come from a simplified version of the one-zone chemical evolution code \texttt{OMEGA} (\citealt{2017ApJ...835..128C}). We adopted a constant star formation rate as a function of time and a closed-box environment, implying no gas exchange between the galaxy and its surrounding environment.  This simplification, although not realistic, was adopted to facilitate the understanding of the decreasing trend of [Eu/Fe] in a GCE context. The top panel of Figure~\ref{fig:knee} shows the production rate of Fe by CC~SNe (blue solid line) and SNe~Ia (red solid line), and the production rate of Eu by NS-NS mergers (black solid and dashed lines), following the two approaches described above.  The interpretation of the residual pink dashed line is discussed in Section~\ref{sec:scenario_5}.

In this simplified framework, because massive stars only live for a few tens of Myr, the Fe production reaches an equilibrium after $\sim$\,40\,Myr.  The production of Fe by SNe~Ia does not show this equilibrium, since SNe~Ia have a wide range of delay times spanning from $\sim$\,100\,Myr to more than $\sim$\,10\,Gyr (e.g., \citealt{2009ApJ...699.2026R}).  This means that the first stellar population formed at the beginning of the simulation will still produce SNe~Ia after $\sim$\,10\,Gyr, which is roughly the lifetime of the simulated galaxy.

When assuming that NS-NS mergers follow the lifetime of massive stars, the production rate of Eu also reaches an equilibrium before the onset of SNe~Ia (black solid line in top panel of Figure~\ref{fig:knee}). Under this assumption, when SNe~Ia start to appear at $\sim$\,100\,Myr, the [Eu/Fe] ratio will bend and start to decrease as a function of time (black solid line in bottom panel of Figure~\ref{fig:knee}).  Indeed, while the production rate of Eu is in equilibrium, the production rate of Fe is increasing, thus progressively reducing the [Eu/Fe] ratio.

On the other hand, when using a long-lasting DTD function in the form of $t^{-1}$, the production rate of Eu no longer reaches an equilibrium (black dashed line in top panel of Figure~\ref{fig:knee}).  In fact, because the adopted DTD function of SNe~Ia is also in the form of $t^{-1}$ (see Section~\ref{sec:obs_GRB}), the production rates of Eu and Fe evolve in a similar way once SNe Ia start to contribute. This results in a nearly constant value for [Eu/Fe] beyond $\sim$\,100\,Myr (black dashed line in bottom panel of Figure~\ref{fig:knee}).  This flat trend, however, is not consistent with the decreasing trend observed in the Galactic disk (see Figure~\ref{fig:Eu_data}).

\subsection{Common Message Sent by Different Studies}
\label{sect_common_message}
The inability of reproducing the decreasing trend of [Eu/Fe] in the Galactic disk using NS-NS mergers with a DTD function in the form of $t^{-1}$ is independent of the complexity of the GCE simulations. Indeed, when assuming constant delay times for NS-NS mergers, all Milky Way simulations reproduce the decreasing trend of [Eu/Fe] (\citealt{2004A&A...416..997A,2014MNRAS.438.2177M,2015A&A...577A.139C,2015MNRAS.452.1970W,2017ApJ...836..230C}).  On the other hand, when assuming a DTD function in the form of $t^{-1}$, all GCE simulations fail to match the decreasing trend of [Eu/Fe] for $-1<$~[Fe/H]~$<0$ (\citealt{2015ApJ...807..115S,2015MNRAS.447..140V,2016ApJ...830...76K,2017ApJ...836..230C,2018arXiv180101141H,2018MNRAS.tmp..585N}). The latter studies represent a wide variety of GCE approaches including simple one-zone models, semi-analytic models of galaxy formation, and cosmological hydrodynamic zoom-in simulations.

\cite{2014MNRAS.438.2177M}, \cite{2015A&A...577A.139C}, and \cite{2017ApJ...836..230C} suggested that GCE predictions at [Fe/H]\,$>$\,$-1$ are unaffected by the choice of the minimum delay time for NS-NS mergers\footnote{These studies have explored delay times within 100\,Myr.} (see Appendix~\ref{app:gce}).  In order to recover the decreasing trend of [Eu/Fe], it seems not sufficient to make NS-NS mergers appear before SNe~Ia, they also need to reach an equilibrium in the production of Eu.  Since [Eu/Fe] and [$\alpha$/Fe] both start to decrease at [Fe/H]\,$\sim$\,$-1$, this equilibrium must be reached within the first few hundreds of Myr.  Although in this section we assumed NS-NS mergers to be the only r-process site, the equilibrium requirement would be the same for any other site.  It can also be applied in the context of multiple r-process sites (see Section~\ref{sec:scenario_5}). We refer to \cite{2018arXiv180101141H} for a discussion on the impact of the minimum delay time of SNe~Ia, as it could help generating a decreasing trend if that minimum delay time is set to 400\,Myr or 1\,Gyr.

\subsection{Confirmation From Our Study}
\label{subsec_GCE_1}
To support the conclusion made in Section~\ref{sect_common_message}, we tested the $t^{-1}$ DTD function using the benchmark Milky Way model of \cite{2014MNRAS.438.2177M}, which is based on the two-infall model originally described in \cite{1997ApJ...477..765C}. We refer to \cite{2019arXiv190102732S} for more details on this new implementation. In this multi-zone framework, which relaxes the instantaneous recycling approximation, the halo and thick disk form on a relatively short timescale (1-2 Gyr) by accretion of primordial gas. This represents the first infall event, while the thin disk forms on a much longer timescale by means of a second independent episode of gas accretion.  The thin disk is assumed to form inside out with a timescale of 7\,Gyr for the solar neighborhood. In this work, we only focus on the thin disk.

\begin{figure}
\center
\includegraphics[width=3.35in]{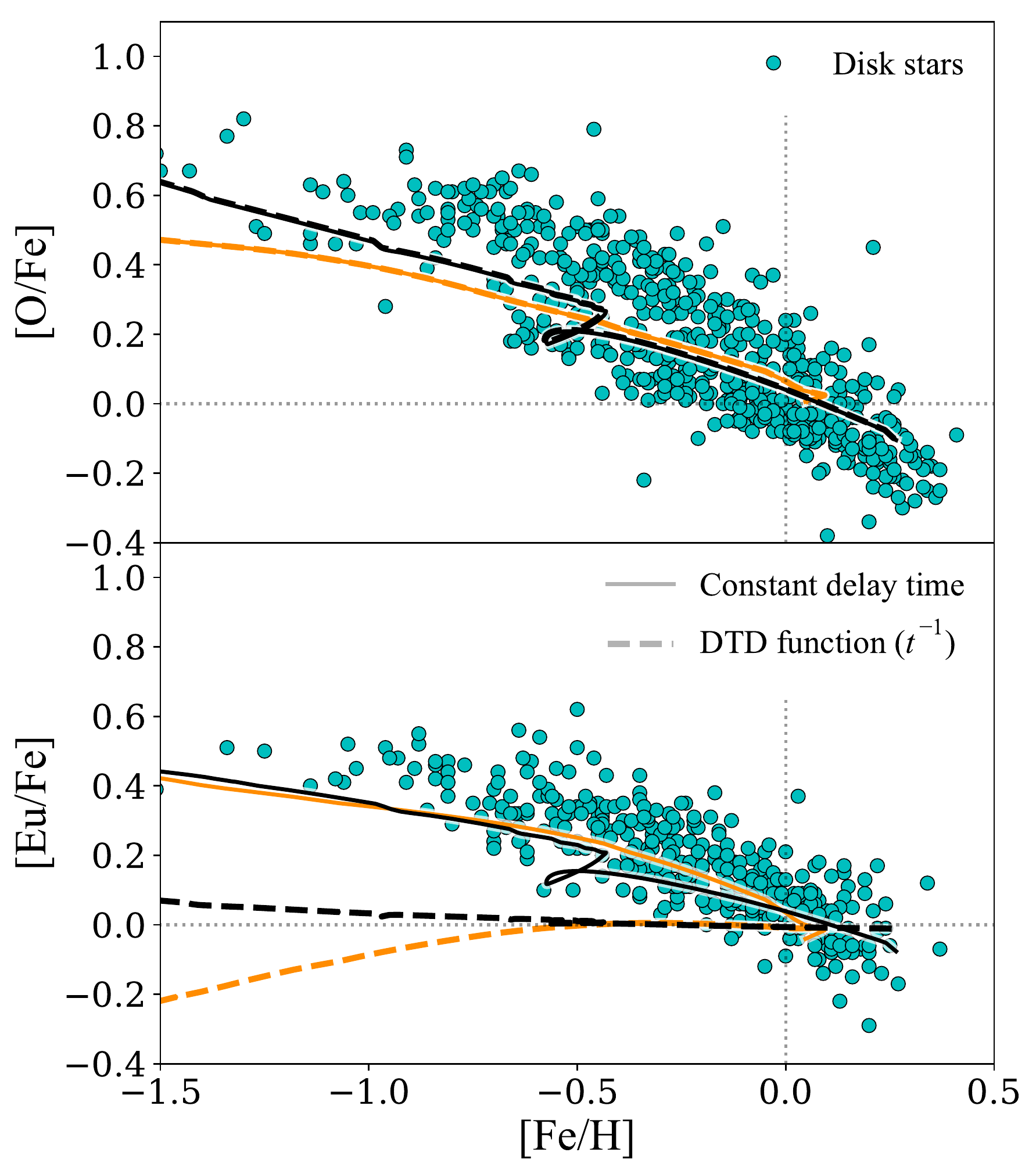}
\caption{Predicted evolution of oxygen  ([O/Fe]) and europium ([Eu/Fe]) abundances as a function of iron abundances ([Fe/H]) in the Galactic disk, using NS-NS mergers and the chemical evolution model of \cite{2017ApJ...835..128C} (\texttt{OMEGA}, orange lines) and \cite{2014MNRAS.438.2177M} (black lines, see Section~\ref{subsec_GCE_1}). The solid line shows the predictions when assuming that NS-NS mergers all occur after a constant delay time of 1\,Myr following the formation of the binary systems, while the dashed line shows the predictions when assuming a delay-time distribution function in the form of $t^{-1}$ to distribute NS-NS mergers as a function of time. Data derived from spectroscopy (cyan dots) are from \cite{2014A&A...562A..71B} for [O/Fe] and from \cite{2016A&A...586A..49B} for [Eu/Fe]. The dotted black lines mark the solar values (\citealt{2009ARA&A..47..481A}).
\label{fig:gce_1}}
\end{figure}

The model predicts at all times the gas fraction and its chemical composition. It takes into account the enrichment from stars of all masses ending their lives as white dwarfs and SNe of all types (II, Ia, Ib, Ic), in addition to the nucleosynthesis occurring in novae and compact binary mergers. The adopted yields are taken from \cite{2010ASSP...16..107K} and \cite{2014MNRAS.437..195D,2014MNRAS.441..582D} for asymptotic giant branch (AGB) and super-AGB stars, from \cite{2013ARA&A..51..457N} for massive stars, from  \cite{1999ApJS..125..439I} for SNe~Ia, from \cite{1998ApJ...494..680J} for novae, and from \cite{korobkin2012} for NS-NS mergers. The SN~Ia rate is computed following the formalism of \cite{2005A&A...441.1055G}. The rate of NS-NS mergers is calculated by convolving its DTD function with the star formation history of the Milky Way, which is generated following a Kennicutt-Schmidt law with a threshold in the gas surface density (\citealt{1998ApJ...498..541K}). The fraction of NS-NS binaries that eventually merge has been tuned to reproduce the current rate derived in \cite{2017ApJ...848L..13A}. 

The bottom panel of Figure~\ref{fig:gce_1} shows the predictions of this model when using a constant delay time (black solid line) and a $t^{-1}$ DTD function (black dashed line) to calculate the rate of NS-NS mergers.  As in previous studies (see Section~\ref{sect_common_message}), the long-lasting DTD function generates a flat trend for [Eu/Fe] and does not reproduce the decreasing trend of the Galactic disk (cyan dots). The little loop at [Fe/H]\,$\sim$\,$-0.5$ is caused by the second infall episode that introduces primordial gas in the galaxy that momentarily dilutes the Fe concentration relative to H. The top panel Figure~\ref{fig:gce_1} shows the evolution of [O/Fe] as a comparison baseline for the decreasing trend of [Eu/Fe].

Throughout this study, we also use the GCE code \texttt{OMEGA} (\citealt{2017ApJ...835..128C}) in order to explore different DTD functions for NS-NS mergers (see Section~\ref{sec:scenario_3}), and to compare with the results of the GCE code of \cite{2014MNRAS.438.2177M} (see orange lines in Figure~\ref{fig:gce_1}). It consists of a classical one-zone model that adopts homogeneous mixing but relaxes the instantaneous recycling approximation. SNe~Ia are distributed in time following a DTD function in the form of $t^{-1}$ that is multiplied by the fraction of white dwarfs (see \citealt{2016ApJ...824...82C} and \citealt{2017arXiv171109172R} for more details.) Yields for low- and intermediate-mass stars, massive stars, SNe~Ia, and NS-NS mergers are taken from \cite{2018MNRAS.480..538R}, \cite{2013ARA&A..51..457N}, \cite{1999ApJS..125..439I}, and \cite{2007PhR...450...97A}, respectively. In this paper, every \texttt{OMEGA} simulations have the same physical setup, only the delay time assumptions of NS-NS mergers are changed. By the end of all \texttt{OMEGA} simulations, the mass of gas within our Milky Way model is $9.5\times10^{9}$\,M$_\odot$, the star formation efficiency is $2.9\times10^{-10}$\,yr$^{-1}$, and the star formation rate is 2\,M$_\odot$\,yr$^{-1}$.

\subsection{Population Synthesis Models}
\label{sec:pop_synth}
Population synthesis models aim to follow the evolution of binary systems involving stars and compact remnants (white dwarfs, neutron stars, and black holes).  The predictions of such models are typically for individual stellar populations where all stars are assumed to form at the same time with the same initial metallicity. In the context of this paper, these models are important because they predict the DTD functions of NS-NS and BH-NS mergers, which are crucial inputs for GCE simulations.

\begin{figure*}
\center
\includegraphics[width=7.in]{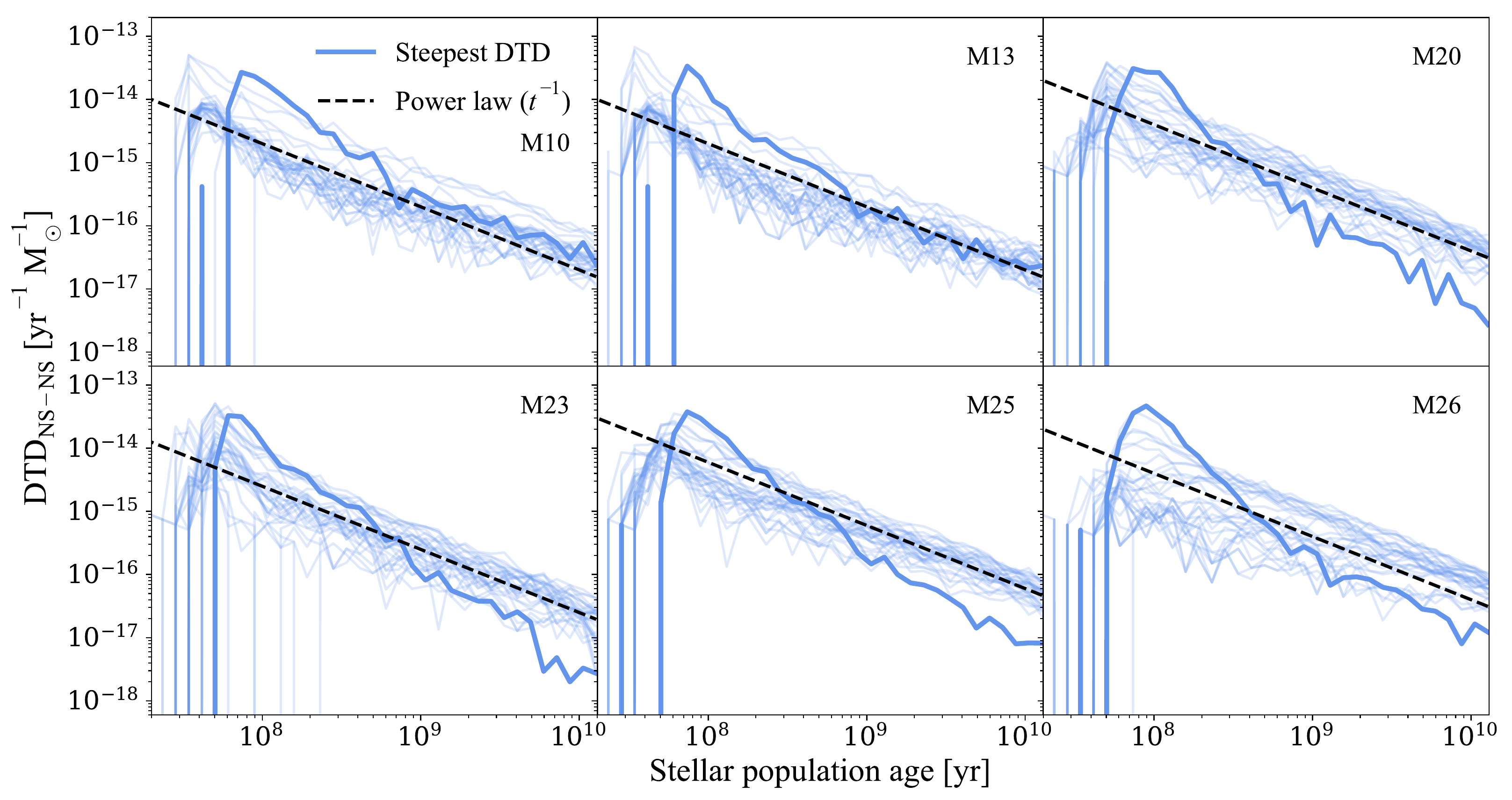}
\caption{Delay-time distribution (DTD) functions of NS-NS mergers predicted by the population synthesis models described in Section~\ref{sec:pop_synth}, for a stellar population normalized to 1\,M$_\odot$. Each panel represents a different set of assumptions regarding the evolution of binary systems. The thin blue lines show the predictions for 32 different initial metallicities.  The thick blue line shows the prediction with the steepest slope (to be used as an argument in Section~\ref{sec:scenario_3}).  As a reference, the dashed black line shows a power-law DTD in the form of $t^{-1}$.
\label{fig:dtd_nsns}}
\end{figure*}

For this study, we use the already-calculated models M10, M13, M20, M23, M25, and M26 (see \citealt{Belczynski2018b}) obtained with the upgraded population synthesis code {\tt StarTrack} \citep{Belczynski2002,Belczynski2008}. These six models aim to provide a range of solutions for the predicted DTD functions that reflects uncertainties in the evolution of binary systems. A short description of the {\tt StarTrack} code and the models is included in Appendix~\ref{app:popsynth}.

Figure~\ref{fig:dtd_nsns} shows the DTD functions predicted by \texttt{StarTrack} for NS-NS mergers. For each model (each panel), the predictions at different metallicities are overall consistent with a simple power law in the form of $t^{-1}$. There are, however, interesting cases showing steeper power-law indexes (thick blue line) that are explored in Section~\ref{sec:scenario_3}.  These results are consistent with other population synthesis studies \citep{2018MNRAS.474.2937C,Kruckow2018,Vigna2018,Giacobbo2018}.

An important physical ingredient driving the slope of the DTD function of NS-NS mergers is the distribution of orbital separations ($a$) of massive stars in binary systems (e.g., \citealt{2017ApJ...846..170T}). According to \cite{2018arXiv181210065B}, an orbital separation distribution in the form of $a^{-1}$ tends to generate NS-NS merger DTD functions in the form of $t^{-1}$, while a distribution in the form of $a^{-3}$ tends to generate steeper NS-NS merger DTD functions in the form of $t^{-1.5}$ (see also Figure~2 in \citealt{2018A&A...615A..91B}).

\begin{figure*}
\center
\includegraphics[width=7.in]{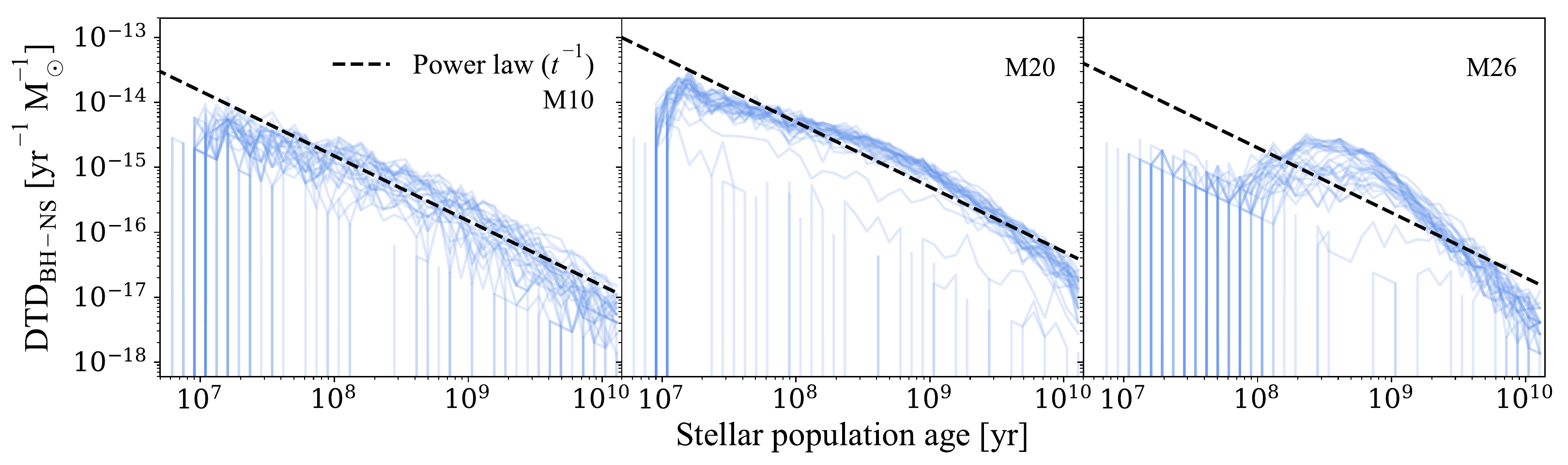}
\caption{Same as in Figure~\ref{fig:dtd_nsns}, but for BH-NS mergers. Predictions showing vertical drops and discontinuities did not have sufficient BH-NS mergers to generate a statistically meaningful DTD function.  Only the M10, M20, and M26 models are shown, as the others did not predict enough BH-NS mergers to generate conclusive DTD functions.
\label{fig:dtd_bhns}}
\end{figure*}

Figure~\ref{fig:dtd_bhns} shows the predictions for BH-NS mergers for the three models that generated the highest number of BH-NS mergers. Relative to NS-NS mergers, BH-NS mergers can appear earlier during the lifetime of a stellar population. But, as described in Section~\ref{sect_common_message}, this fact alone is unlikely to help generating a decreasing trend for [Eu/Fe] in chemical evolution simulations. The shape of their DTD functions is less in agreement with a power law in the form of $t^{-1}$, as there is typically a bump in between 0.1 and 1\,Gyr.  Compared to a $t^{-1}$ power law, this implies a flatter power law before 1\,Gyr followed by a steeper power law after 1\,Gyr, at least for the M20 and M26 models.

\section{R-Process Nucleosynthesis Calculations}
\label{sec:nucleo}
Here we discuss the connections of the GCE results above to theoretical r-process calculations. We give a more detailed review of the nucleosynthesis of r-process elements in different types of environments in Appendix~\ref{sec_nucleo}. R-process nucleosynthesis calculations are usually based on hydrodynamic simulations of scenarios that can provide very neutron-rich conditions. Among those, the most promising are currently compact binary mergers and magneto-rotational (MR) supernovae, a rare class of CC~SNe. Other possible scenarios include magnetized winds from proto-neutron stars (\citealt{2003ApJ...585L..33T,2007ApJ...659..561M,2008ApJ...676.1130M,2014MNRAS.444.3537V,2018MNRAS.476.5502T}), pressure-driven outflows \citep{fryer06}, accretion disk winds from collapsars (\citealt{2018arXiv181000098S}), and the jets generated by a neutron star that accretes matter from its giant companion star \citep{papish2015,2017ApJ...851...95S,2018arXiv181003889G}.
 
 The evolution of the mass ejected depends on the evolution of density and temperature as a function of time, and on the initial electron fraction ($Y_e$). The latter serves as a measurement of the neutron fraction in the environment as it is defined as the proton-to-baryon ration, $Y_e=Y_p/(Y_p+Y_n)$. If the $Y_e$ is low ($Y_e \lesssim 0.25$), the neutron fraction ($Y_n$) will be high, and more neutrons will be available to be captured by seed nuclei, i.e. the neutron-to-seed ration ($Y_n/Y_\mathrm{seed}$) is high. Typically, the $Y_e$ value is taken when the temperature drops below $\sim$\,10~GK. The evolution of density and temperature is often categorized by the entropy and the expansion timescale of the ejecta, which are used as nucleosynthesis parameters. 
 
\subsection{Robustness of the Main r-Process}
\label{sec:nucleo_robust}
An individual simulation of an environment can provide very different conditions for the ejecta.  When several simulations are considered, the variations are even larger. For example, with NS-NS mergers, one can study different possibilities by varying the masses of both neutron stars, the equation of state, and other input physics. We say that the r-process nucleosynthesis is robust when the relative abundance pattern is insensitive to such variations and is always similar between the second and third r-process peak. If the initial neutron-to-seed ratio is high enough ($Y_{n}/Y_\mathrm{seed} > 150$), the r-process path runs into isotopes that are prone to fission, producing two (or more) fission fragments that can in turn continue to capture neutrons. This so-called ``fission cycling'' guarantees that the final yields produce a robust and reproducible abundance pattern, independently of the exact hydrodynamical conditions \citep{beun2006,goriely2011,roberts2011,korobkin2012,rosswog2013,eichler2015,mendoza2015}.
\\
\\
\subsection{Neutron Star Mergers vs Supernovae}
\label{sec:mergers_vs_sn}
Since the uncertainties in nuclear properties and reaction rates involved in the r-process are still very large, model compositions for ejecta of different r-process sites should be considered qualitative predictions rather than precise results. Nevertheless, some key differences can be expected in the final abundance patterns of NS-NS mergers and MR~SNe, if the conditions in the ejecta are reasonably different.

Hydrodynamic simulations suggest that the dynamical ejecta of NS-NS mergers are generally more neutron-rich than those of MR~SNe, even if the entropy range is similar. This directly translates to higher initial neutron-to-seed ratios and, as a consequence, to a larger fraction of the ejecta where the third peak and the actinides region can be reached. While a large fraction of NS-NS mergers ejecta co-produce the second and third r-process peaks, MR~SN ejecta contain regions where the neutron-to-seed ratio is not large enough to produce third-peak elements, leading to the build-up of a distinct second r-process peak. As a result, the abundance ratio between the third and second peak for the total ejecta should be larger in dynamical ejecta of NS-NS mergers than in MR~SNe. Varying contributions from disk and neutrino-driven wind ejecta in NS-NS mergers could unfortunately wash out this distinction.

Another observable could be the shape of the second peak itself. As described above, in MR~SN ejecta, the predominant contribution in the build-up of the second peak comes from regions with moderate $Y_e$ where the r-process flow is stopped at the shell closure at $N = 82$. In that case, the final shape of the second peak is mostly determined by $\beta$-decays and $\beta$-delayed neutron emissions. On the other hand, the neutron-rich NS-NS merger ejecta can produce a considerable amount of actinides which fission (and $\alpha$-decay) on timescales that are similar to or longer than the r-process duration. Most nuclear mass models predict the majority of fissioning nuclei in the mass range $240 < A < 280$, which leads to fission fragments at or around the second peak. The production of fission fragments after the end of the r-process implies that the second peak in this case is shaped by the fission fragment distribution and their subsequent $\beta$-decays on top of the abundances already present at the time of r-process freeze-out\footnote{Freeze-out is defined as the moment when the neutrons are almost exhausted and thus $Y_n/Y_\mathrm{seed}\leq 1$. This coincides with time when $\beta$-decays become faster than neutron captures and can identified with the corresponding time scales: $\tau_\beta < \tau_{(n,\gamma)}$}.

\begin{figure}
\center
\includegraphics[width=3.35in]{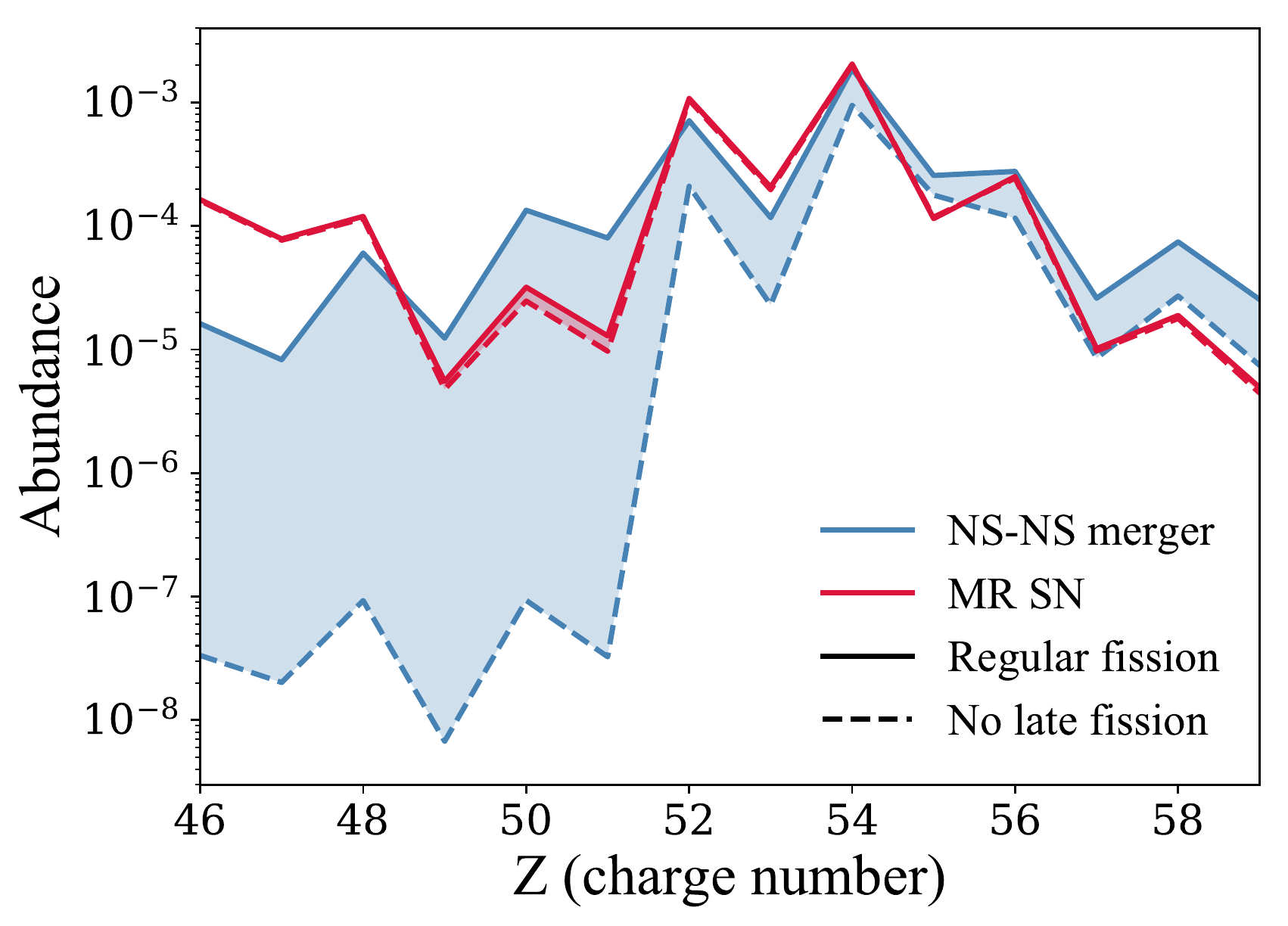}
\caption{Predicted abundances of the second r-process peak for a neutron star (NS-NS) merger calculation (blue) and a magneto-rotational supernova (MR~SN) scenario (red), assuming either a continuous fission fragment production (``regular fission'') or a setup where fission fragment production is disabled after the r-process freeze-out (``no late fission''). The shaded regions represent the fission contribution to the final abundances of the second peak. \label{fig:secondpeak}}
\end{figure}

To illustrate this, we have performed r-process calculations based on the NS-NS merger simulation of \cite{rosswog2013} involving two neutron stars of 1.4~M$_{\odot}$ each, and based on the MR~SN model of \cite{winteler2012} (see Figure~\ref{fig:secondpeak}). Both sets of calculations were performed with fully enabled fission reactions (blue and red solid lines), as well as with a modified setup disabling fission from the moment when $Y_n/Y_\mathrm{seed} \leq 1$ onwards. The dashed lines in Figure~\ref{fig:secondpeak} represent the $\beta$-decayed second peak material already present at the r-process freeze-out without any later fission contribution. The differences in the two approaches therefore arise solely from fission fragments after the r-process freeze-out. 

We should point out, however, that the distributions of fission fragments hold at least the same uncertainties as the nuclear reaction rates in the r-process. Our calculations shown in Figure~\ref{fig:secondpeak} have been performed using the ABLA07 fission fragment distribution model \citep{kelic2009}. Other fission models predict different distributions \citep{eichler2015, goriely2015}. Nevertheless, the different mechanisms forming the second r-process peak should result in different peak shapes, as long as fission produces fragments in the second peak. Following this argumentation, metal-poor stars with larger actinide abundances should also exhibit second peak abundances reflecting the fission fragment production more closely than stars with a lower actinide content.  

\section{R-Process Sites and their Challenges}
\label{sec:challenges}
This section highlights the agreements and inconsistencies that emerge when assuming that current models of NS-NS mergers or MR~SNe are the only source of r-process elements in galaxies. The results of this investigation are summarized in Table~\ref{tab_summary}. In particular, we find that NS-NS mergers alone cannot be consistent with all the observational constraints listed in Section~\ref{sec:obs}.  In Section~\ref{sec:scenario_5}, we discuss the possibility of multiple sites for the heavy r-process element production.

\subsection{Neutron Star Mergers with no Delay-Time Distribution}
\label{sec:scenario_1}
As described in Section~\ref{sect_NSNS_GCE}, delaying the occurrence of NS-NS mergers by a constant coalescence timescale following the lifetime of massive stars (i.e., not taking into account a DTD function) allows GCE simulations to reproduce the decreasing trend of [Eu/Fe] in the Galactic disk. If NS-NS mergers are the dominant r-process site, it guarantees a robust r-process pattern for elements between the second and third r-process peaks (see Sections~\ref{sec:obs_robustness} and Appendix~\ref{sec_nucleo_nsm}). In addition, they eject a sufficiently large amount of material per merger event to explain the large scatter of [Eu/Fe] ratios in the metal-poor stars in the Galactic halo (see Section~\ref{sec:obs_scatter}).

However, because this scenario does not use a DTD function for the coalescence times of NS-NS mergers, but rather adopts a maximum coalescence time of $\lesssim$~100\,Myr, it is incompatible with several observations.  In particular, it cannot explain the detection of short GRBs in early-type galaxies (see Section~\ref{sec:obs_GRB_host}) and the long-lasting DTD functions predicted by population synthesis models (see Section~\ref{sec:pop_synth}) and expected theoretically from the wide variety of orbital parameters in binary systems (\citealt{2015IJMPD..2430012R}). Furthermore, a constant coalescence time of $\lesssim$\,100\,Myr is incompatible with the seven known NS-NS binaries with estimated coalescence times ranging from 46 to 2730\,Myr (see \citealt{2017ApJ...846..170T} and Section~\ref{sec:obs_GRB_DTD}). Finally, this scenario does not allow the gravitational wave event GW170817 to be detected in an early-type galaxy.

\subsection{Neutron Star Mergers with a Delay-Time Distribution in the Form of $t^{-1}$}
\label{sec:scenario_2}
A DTD function in the form of $t^{-1}$ for NS-NS mergers is consistent with the detection of short GRBs (see Section~\ref{sec:obs_GRB_DTD}) and the predictions of population synthesis models (see Section~\ref{sec:pop_synth}).  It is also consistent with the fact that SNe~Ia and short GRBs are both detected in similar proportions in early-type galaxies (see Section~\ref{sec:obs_GRB}), given that most studies agree that SNe~Ia do follow a DTD function in the form of $t^{-1}$ as well (but see \citealt{2017ApJ...834...15H}).  Many GCE simulations have explored the role of NS-NS mergers on the chemical evolution of Eu in the Milky Way using this canonical DTD function. However, as described in Section~\ref{sect_NSNS_GCE}, this scenario does not allow to reproduce the decreasing trend of [Eu/Fe] in the Galactic disk between [Fe/H]~$=-1$ and~$0$, assuming that NS-NS mergers are the only source of heavy r-process elements.

A potential solution to this discrepancy is to claim that short GRB observations suffer from a lack of statistics, and that more detections should increase the fraction of short GRBs in late-type star-forming galaxies. This would imply a steeper DTD function for NS-NS mergers (see Section~\ref{sec:scenario_3}).  Another solution is to assume a flatter DTD function for SNe~Ia.  But this goes in the opposite direction of recent observations suggesting that, if the DTD function of SNe~Ia is different than $t^{-1}$, it should be steeper (\citealt{2017ApJ...834...15H}).

\subsection{Neutron Star Mergers with Our Exploratory Delay-Time Distributions}
\label{sec:scenario_3}
\begin{figure}
\begin{tabular}{c}
\hspace*{-0.55cm}
\includegraphics[width=3.35in]{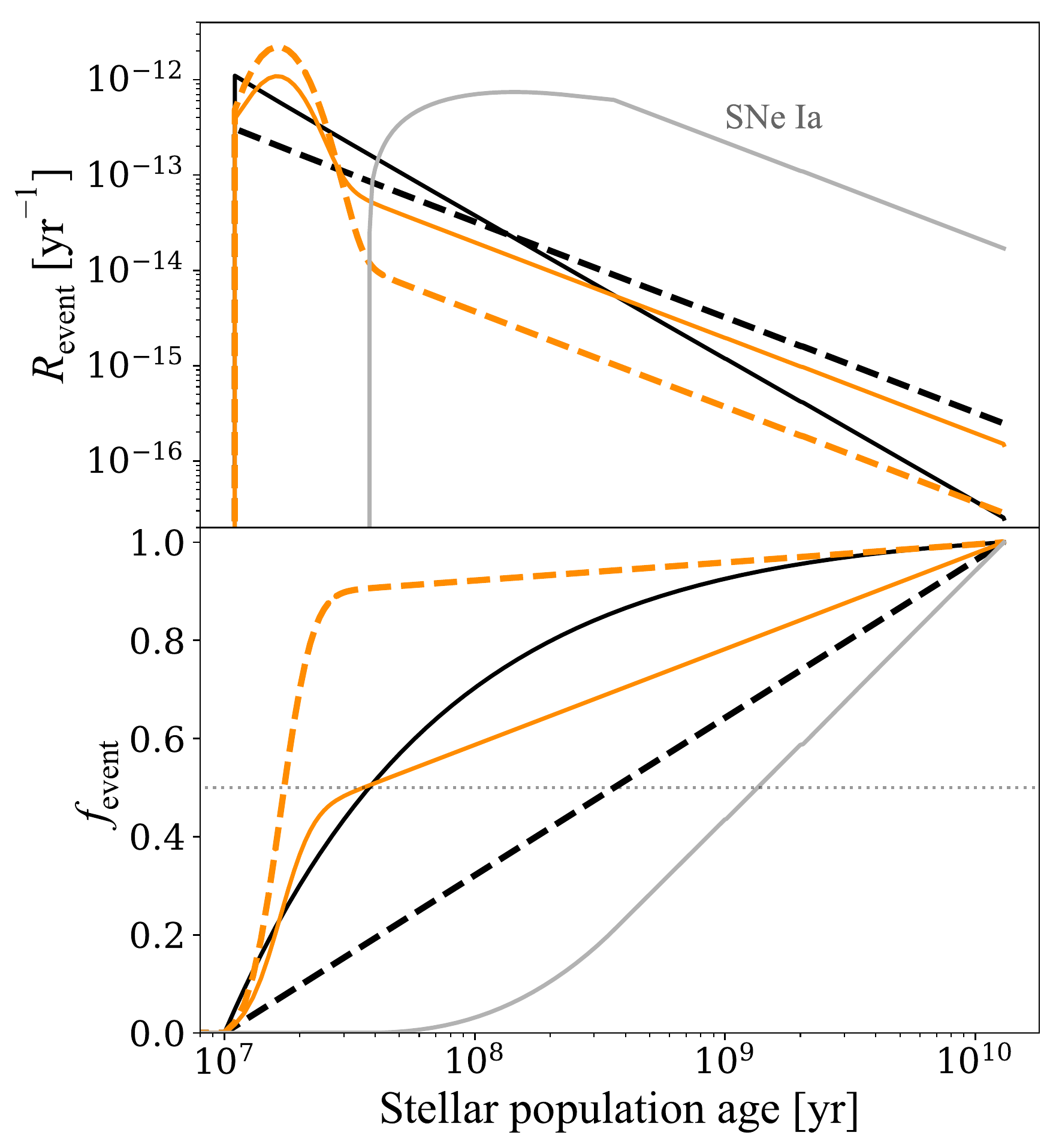} \\
\includegraphics[width=3.35in]{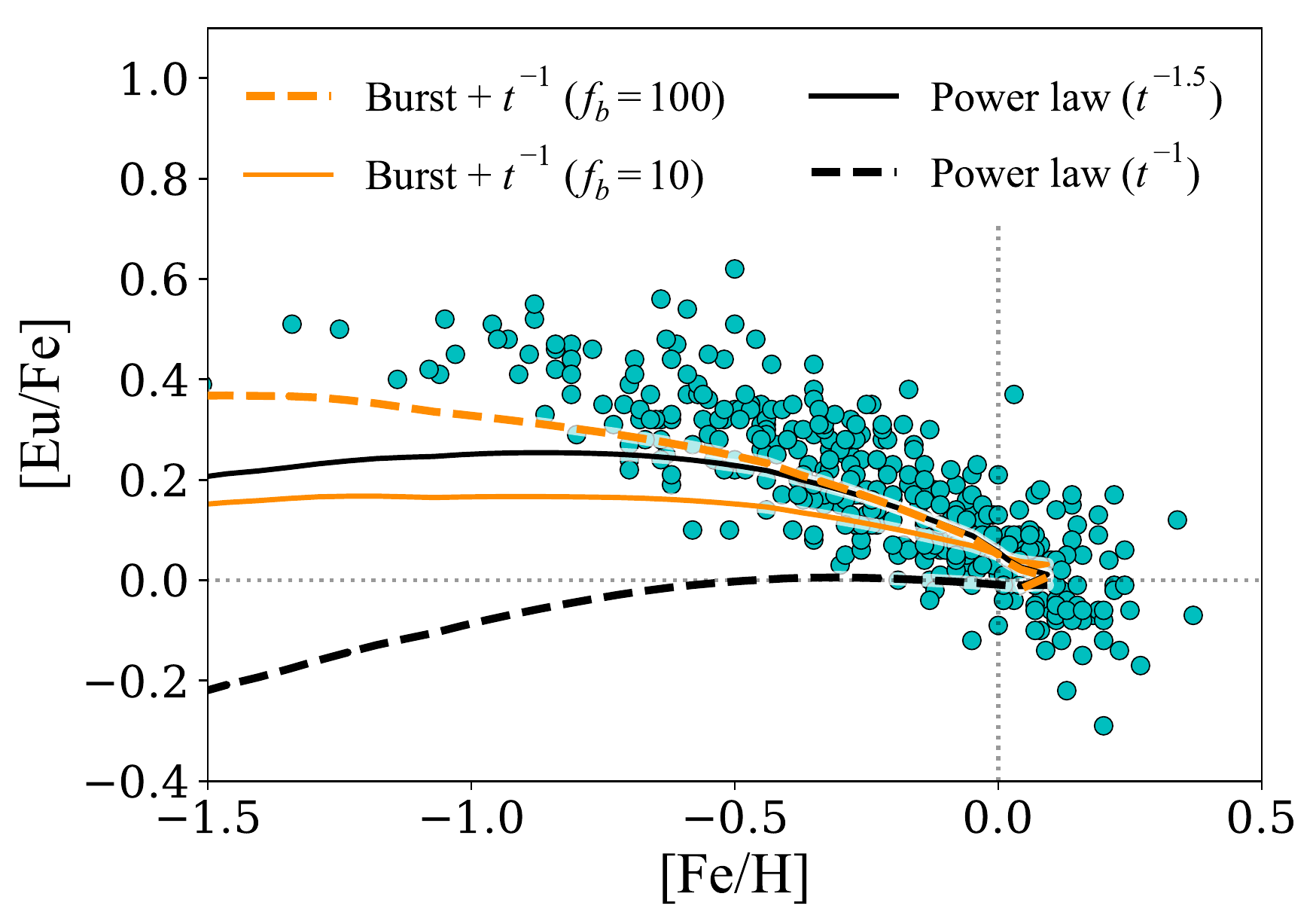}
\end{tabular}
\caption{Impact of using different delay-time distribution (DTD) functions for NS-NS mergers on the predicted chemical evolution of Eu. \textbf{Top panel}: Four different DTD functions (black and orange lines) in a simple stellar population of 1\,M$_\odot$. Once integrated, all four DTD functions produce the same number of NS-NS mergers.  The grey solid line represents the DTD function used for SNe~Ia. \textbf{Middle panel}: Cumulated fraction of NS-NS mergers and SNe~Ia as a function of time in a stellar population. The dotted horizontal line marks the moment where 50\,\% of the events have occurred. \textbf{Bottom panel}: Predicted chemical evolution of [Eu/Fe] as a function of [Fe/H] using the different DTD functions presented in the top panel. For DTD functions with bursts, $f_b$ is enhancement factor of the burst relative to the value given by the background power-law distribution. Cyan dots are stellar abundances data for disk stars taken from \cite{2016A&A...586A..49B}. The dotted black lines mark the solar values (\citealt{2009ARA&A..47..481A}).} \label{fig:mod_DTD}
\end{figure}

As discussed in the last section, using a DTD function in the form of $t^{-1}$ for NS-NS mergers in chemical evolution simulations does not allow to reproduce the decreasing trend of [Eu/Fe] in the Galactic disk.  In this section, we test alternative and exploratory DTD functions, which are shown in the top panel of Figure~\ref{fig:mod_DTD} (see also \citealt{2019arXiv190102732S}). The idea is to recover the decreasing trend of [Eu/Fe] by concentrating the bulk of NS-NS mergers at early times, before the onset of SNe~Ia, while still allowing NS-NS mergers to have a wide range of coalescence timescales (up to $\sim$\,10\,Gyr). The predictions shown in Figure~\ref{fig:mod_DTD} have been computed with the simple stellar population code \texttt{SYGMA} (\citealt{2017arXiv171109172R}) and the galactic chemical evolution code \texttt{OMEGA} (\citealt{2017ApJ...835..128C}). Both codes are part of the open-source NuPyCEE package\footnote{\url{https://github.com/NuGrid/NuPyCEE}}.

As shown in the bottom panel of Figure~\ref{fig:mod_DTD}, it is possible to recover a decreasing trend for [Eu/Fe] using a steep power law in the form of $t^{-1.5}$ (black solid line, see also \citealt{2017ApJ...836..230C} and \citealt{2018arXiv180101141H}). This could be in agreement with the $t^{-1.5}$ DTD function derived by \cite{2015JHEAp...7...73D} for short GRBs, although the recent analysis of \cite{2017ApJ...848L..23F} rather points toward a function in the form of $t^{-1}$. We note that some of our population synthesis models do predict DTD functions steeper than $t^{-1}$ (see thick blue line in Figure~\ref{fig:dtd_nsns}), although the majority are consistent with a $t^{-1}$ distribution.  We note that the presence of a third object interacting with a NS-NS binary could potentially help to concentrate NS-NS mergers at shorter coalescence timescales (\citealt{2018PASA...35...17B}).

\begin{figure}
\includegraphics[width=3.35in]{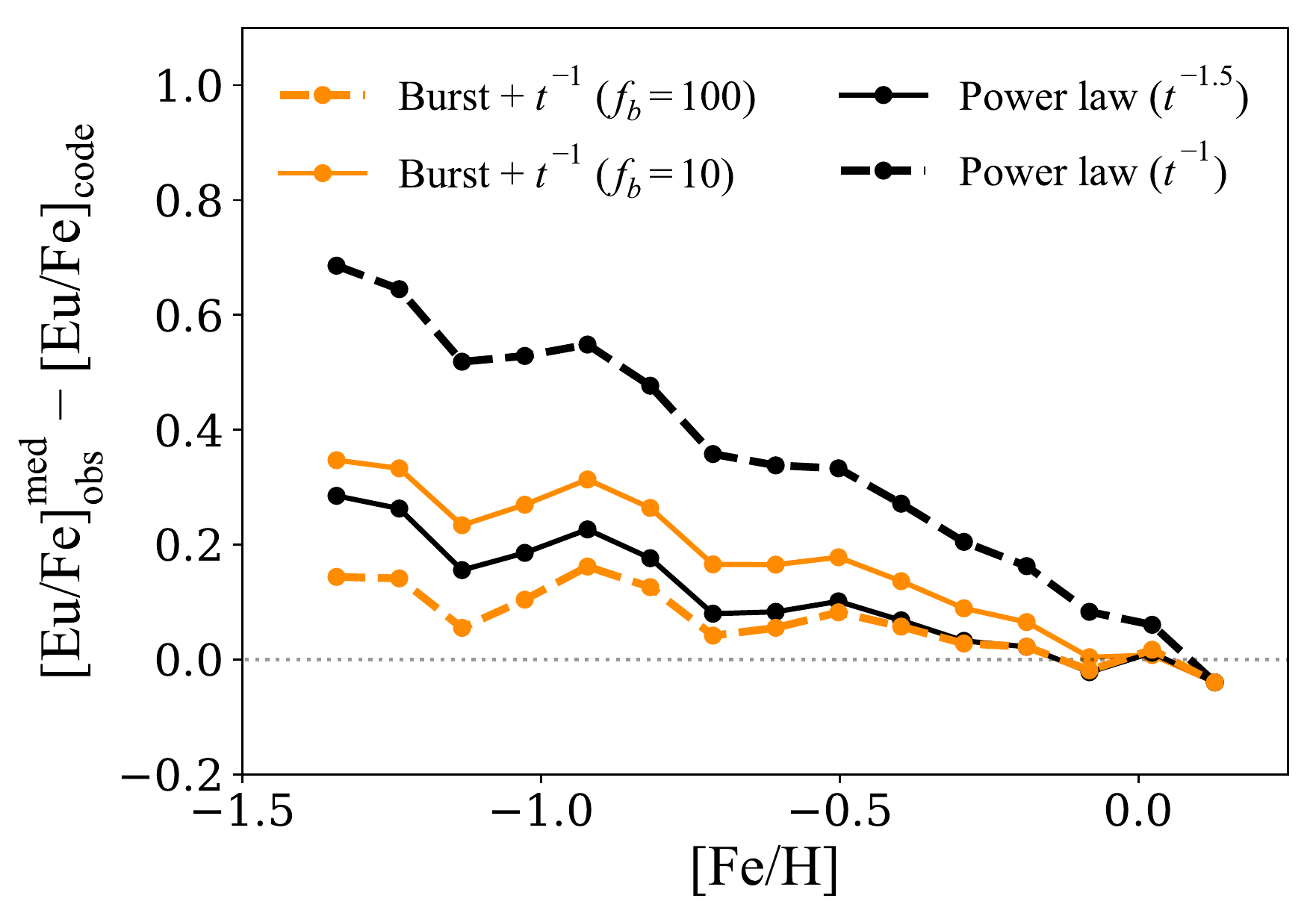}
\caption{Difference between the [Eu/Fe] trends observed in the Galactic disk and predicted using different delay-time distribution (DTD) functions for NS-NS mergers, as presented in Figure~\ref{fig:mod_DTD}. For the comparison, we split the [Fe/H] axis into bins with constant intervals of 0.105. The dots represent the central value of those bins. For each bin, we calculated the median value of all data present in that bin, and subtracted the predictions made by our code (see Section~\ref{sec:scenario_3}). We interpolated our predictions between the timesteps in order to recover the values at the central [Fe/H] bin points.} \label{fig:slope}
\end{figure}

Alternatively, if one assumes a strong burst of mergers before the onset of SNe~Ia, followed by a power law in the form of $t^{-1}$, the decreasing trend of [Eu/Fe] can also be recovered (see orange lines in Figure~\ref{fig:mod_DTD}). This type of DTD function for NS-NS mergers has been seen in the population synthesis predictions of \citealt{2012ApJ...759...52D}, as shown in the lower-left panel of their Figure~8. When using the DTD functions predicted by \citealt{2012ApJ...759...52D}, the decreasing trend of [Eu/Fe] has been recovered with the GCE code \texttt{OMEGA} (\citealt{2017ApJ...835..128C}), as shown with the black solid line in Figure~5 of \cite{2017MNRAS.469.4378M}.

Figure~\ref{fig:slope} shows the difference between the observed chemical evolution trend of [Eu/Fe] and the trends predicted using the different DTD functions explored in Figure~\ref{fig:mod_DTD}. The steep power law ($t^{-1.5}$) and the DTD with the strongest burst ($f_b=100$) provide the best fit and generate trends that are consistent within 25\,\% with the median trend observed at [Fe/H]$\,>-0.75$. Overall, the more the bulk of Eu is released at early time, the better will be the fit (see middle panel of Figure~\ref{fig:mod_DTD}).

With the DTD functions explored in this section, more than half of all NS-NS mergers produced by a stellar population will have occurred by $\sim$~100\,Myr (see middle panel of Figure~\ref{fig:mod_DTD}). This is because in order to generate a decreasing trend with chemical evolution models, the bulk of Eu must be released before the onset of the Fe production by SNe~Ia.  However, this significant difference between NS-NS mergers and SNe~Ia is currently inconsistent with their similar fraction of occurrence in early-type galaxies (see Section~\ref{sec:obs_GRB_host}). Furthermore, assuming the DTD functions explored in this section are representative of the real ones, detecting the first gravitational wave signal of a NS-NS merger in an early-type galaxy becomes even more improbable, although not impossible (see Section~\ref{sec:obs_grav_host}).

\subsection{Rare Classes of Core-Collapse Supernovae}
\label{sec:scenario_4}
Rare classes of CC~SNe, such as MR~SNe, would naturally explain how Eu can be present in the atmosphere of the most metal-poor stars (\citealt{2004A&A...416..997A,2014MNRAS.438.2177M,2015A&A...577A.139C,2015MNRAS.452.1970W,2017ARNPS..67..253T,2019MNRAS.483.5123H}). In addition, because the lifetime of massive stars is short relative to the production timescale of Fe in SNe~Ia, using CC~SNe in GCE simulations allows to reproduce the decreasing trend of [Eu/Fe] in the Milky Way between [Fe/H]~$=-1$ and 0 (see Section~\ref{sect_NSNS_GCE}).  

But hydrodynamic simulations cannot guarantee that MR~SNe will systematically produce the conditions for a robust r-process pattern between the second and third peak (see Appendix~\ref{sec:cc_sne_robust}).  While some parts of the ejecta in SNe simulations can synthesize the second and third r-process peaks, there are several other parts (other trajectories) that will mostly produce the first and second peaks only. When summed together, it is therefore not guaranteed that the ratio between the second and the third r-process peaks will always be similar from one SN to another. If we assume that rare CC~SNe are the only source of r-process elements, there is then a potential tension with the robust main r-process abundance pattern observed in r-process enhanced metal-poor stars (see Section~\ref{sec:obs_robustness}).  But on the other hand, those SNe could explain the diversity of abundances for the first-peak elements (see Section~\ref{sec:obs_met_poor} and \citealt{2010ApJ...724..975R}).

This scenario could also be in tension with the high NS-NS merger rate established by LIGO/Virgo (see Section~\ref{sec:obs_grav}). Once the rate and ejecta of NS-NS mergers will be better constrained, we will be in a better situation to quantify how much room there is for additional r-process sites besides NS-NS mergers.

\begin{center}
\begin{deluxetable*}{lcccc}
\tablewidth{0pc}
\tablecaption{List of agreements and inconsistencies when assuming that current models of NS-NS mergers and MR~SNe are the only r-process site for elements in between the second and third peaks. Three assumptions are made for the delay-time distribution (DTD) function of NS-NS mergers: 1~-- follow the lifetime of massive stars with short constant delay times (without a DTD function), 2~-- DTD function in the form of $t^{-1}$, 3~-- modified DTD functions, either a steeper $t^{-1.5}$ power law or a $t^{-1}$ power law combined with a prompt burst of mergers before the onset of SNe~Ia.  For each observational constraint, a \textit{Yes} or a \textit{No} highlights whether or not the selected r-process site is consistent with the constraint. A \textit{Maybe} means that the situation is unclear. We refer to Section~\ref{sec:scenario_5} for a discussion on the possible contribution of multiple sites on the total budget of r-process elements in the Milky Way. This table reflects our current state of understanding.  Further observations and/or improved models could change the interpretation shown in this table.
\label{tab_summary}}
\tablehead{ \multirow{3}{*}{Observational Constraints} & \multicolumn{3}{c}{NS-NS Mergers} & {MR~SNe} \\
& \colhead{No DTD} & \colhead{$t^{-1}$ DTD} & \colhead{Modified DTD} & \colhead{ } \\
& \colhead{(Section~\ref{sec:scenario_1})} & \colhead{(Section~\ref{sec:scenario_2})} & \colhead{(Section~\ref{sec:scenario_3})} & \colhead{(Section~\ref{sec:scenario_4})}}
\startdata
Production of a robust main r-process pattern (Section~\ref{sec:obs_robustness}) & Yes & Yes & Yes & Maybe \\
Possibility of producing actinides elements (Section~\ref{sec:obs_robustness}) & Yes & Yes & Yes & Yes \\
Large scatter of [Eu/Fe] in metal-poor stars (Section~\ref{sec:obs_scatter}) & Yes & Yes & Yes & Maybe \\
Stars with [Eu/Fe$]>0.3$ at $[$Fe/H$]\lesssim-2$ (Figure~\ref{fig:Eu_data}) & Maybe & Maybe & Maybe & Yes \\
Decreasing trend of [Eu/Fe] at $[$Fe/H$]>-1$ (Section~\ref{sec:obs_trends}) & Yes & No & Yes & Yes \\
Fraction of short GRBs in early-type galaxies (Section~\ref{sec:obs_GRB}) & No & Yes & No & -- \\
LIGO/Virgo local NS-NS merger rate density (Section~\ref{sec:obs_grav_kilo}) & Yes & Yes & Yes & Maybe \\
\hline 
Probability of detecting GW170817 in S0 galaxy (Section~\ref{sec:obs_grav_host}) & Zero & Low & Very low & -- \\
\enddata
\end{deluxetable*}
\end{center}

\subsection{Multiple r-Process Sites Scenario}
\label{sec:scenario_5}
To summarize the previous sections, neither NS-NS mergers or MR~SNe can  reproduce on their own all the observational constraints simultaneously. One solution to this tension is to assume that some of the observational evidence are biased in some way, and that future observations will modify our interpretation of these constraints. But if the current observations are representative, then there are challenges to be solved. Assuming that future gravitational wave and kilonova detections will prove that NS-NS merger do play an important role in the r-process inventory of the Milky Way, something must be added in chemical evolution simulations in order to recover the decreasing trend of [Eu/Fe] in the Galactic disk. 

A description of how to reproduce such a decreasing trend was presented in Section~\ref{sect_knee}. According to this analysis, the production rate of Eu needs to reach an equilibrium before the onset of SNe~Ia. But the production rate of Eu from NS-NS mergers do not reach such an equilibrium, given the long-lasting nature of their DTD function (see black dashed line in top panel of Figure~\ref{fig:knee}). The black solid line in the top panel of that same figure represents the production rate that is required to generate a decreasing trend, as shown in the bottom panel. But this rate does not have to come from only one site. Indeed, by subtracting the production rate of NS-NS mergers (black dashed line) from the required rate (black solid line), one ends up with a residual production rate, highlighted as a pink dashed line in Figure~\ref{fig:knee}. This residual line represents the production rate that needs to be added to the production rate of NS-NS mergers to recover the decreasing trend of [Eu/Fe].

Assuming that NS-NS mergers do contribute to the chemical evolution of Eu in the Milky Way, this suggests that there should be an additional production site of Eu active in the early universe, at low metallicity (see also Sk{\'u}lad{\'o}ttir et al.~in preparation). To be consistent with chemical evolution studies, this hypothetical extra site should fade away at later times, at higher metallicity.  According to Figure~\ref{fig:knee}, NS-NS mergers should become the dominant r-process site roughly when SNe~Ia start to bend the [Eu/Fe] trend, which roughly corresponds to [Fe/H]\,$\sim$\,$-1$ (e.g., \citealt{2016A&A...586A..49B,2018MNRAS.tmp.1218B}).

By integrating the residual and NS-NS merger lines (dashed lines in Figure~\ref{fig:knee}), about 10\,\% of the current Eu production budget could come from this extra site. With this simple approach, we estimate that this site could have produced 60\,\% and 40\,\% of the Eu during the first 100\,Myr and 1\,Gyr of Galactic evolution, respectively. These estimates are based on the assumptions that NS-NS mergers have a DTD function in the form of $t^{-1}$, and that the overall production rate needed to fit the evolution of [Eu/Fe] is similar to the rate of SNe from massive stars. Furthermore, our estimates depend on the minimum delay time of SNe~Ia and are based on the predictions made by a simple toy model that assumes a constant star formation history. To summarize, our estimates are first order approximations and should serve as a motivation for further studies, rather than be taken for solid predictions.

MR~SNe could provide this extra site of Eu production.  They require a strong magnetic field to synthesize heavy r-process elements (see also, e.g., \citealt{2018MNRAS.476.5502T}), and that magnetic field can be amplified by stellar rotation (see Appendix~\ref{sec:cc_sne_robust}).  Because stars are expected to rotate faster at low metallicity (e.g., \citealt{1999A&A...346..459M,2001A&A...373..555M,2007A&A...462..683M}), it is possible that MR~SNe would preferentially produce the heavy r-process elements in the early universe, while they would mostly produce lighter r-process elements (lighter than Eu) at higher metallicity. In addition, because the nucleosynthesis in MR~SNe is not expected to always reach the second and third r-process peak, they could explain the patterns of limited r-process metal-poor stars (see Section~\ref{sec:obs_limited_r} and \citealt{Hansen2014,Spite2018}), a pattern that is unlikely to occur when summing all ejecta components of a NS-NS merger.

BH-NS mergers could also be this extra site, as long as their contribution peaks at low metallicity. However, as shown in Figure~\ref{fig:dtd_bhns}, the DTD function of BH-NS mergers might not be steep enough to generate a significant boost of Eu before the onset of SNe~Ia. More investigation is needed to validate this scenario. BH-NS mergers will hopefully be discovered in the next observing run of LIGO/Virgo. We stress that any other site that, for some reason, has a peak of Eu production in the early universe is a viable candidate.

We note that previous chemical evolution studies also suggested that SNe alongside with NS-NS mergers should contribute to the chemical evolution of Eu  (\citealt{2004A&A...416..997A,2014MNRAS.438.2177M,2015A&A...577A.139C,2015MNRAS.452.1970W,2019MNRAS.483.5123H}). Although complementary to our work, this was motivated by the difficulty to inject r-process elements early enough to explain the Eu abundances in metal-poor stars (see also \citealt{2018arXiv181202779S}).  Here we suggest the possible contribution of an extra site to solve the problem of reproducing the decreasing trend of [Eu/Fe] in the Galactic disk, a cleaner feature that is the result of $\sim$\,12\,Gyr of chemical evolution in a less stochastic environment.  In addition, our scenario strictly requires that the contribution of this extra r-process site fades away as a function of time and metallicity (pink line in Figure~\ref{fig:Eu_data}). This scenario has also been worked out in parallel by \cite{2018arXiv181000098S}, who suggested that a metallicity-dependent second r-process site (the collapsars), combined with NS-NS mergers, can help to recover the decreasing trend of [Eu/Fe].

As described in Section~\ref{sec:pop_synth}, tighter orbital separations in binary systems can lead to steeper DTD functions for NS-NS mergers. If one assumes that the orbital separation distribution of massive stars was steeper at low metallicity, one could theoretically predict a burst of NS-NS mergers in the early universe, while still recovering a DTD function in the form of $t^{-1}$ at current time (at low redshift). This idea is explored in \cite{2019arXiv190102732S} and could provide a potential solution for the inconsistencies highlighted in Table~\ref{tab_summary}, even with NS-NS mergers only. Furthermore, \cite{2019arXiv190109938S} suggested that the decreasing trend problem for [Eu/Fe] could potentially be solved by accounting for a multi-phase interstellar medium where NS-NS would preferentially deposit their ejecta in a cold gas component, instead of a hot gas component in which core-collapse supernovae deposit their ejecta. This represents a motivation for future studies to better understand the mixing process of r-process elements in the early universe.

\section{Future Studies}
\label{sec:disc}
In this section, we discuss the current state of the hydrodynamic simulations used to calculate r-process nucleosynthesis in compact binary mergers and CC~SNe, and provide guidance for future work to address the tensions highlighted in this paper.

\subsection{Simulations of Compact Binary Mergers}
\label{sec:hydro_sim_cbm}
Most hydrodynamic simulations of compact binary mergers agree on the amount of dynamical mass ejected and on the impact of the mass asymmetry between the two merging objects (see, e.g., Table~1 in \citealt{2018arXiv180504637H}). However, there are disagreements regarding the properties of the ejecta.

Neutrino interactions can change the electron fraction of the ejecta, and therefore the r-process nucleosynthesis (e.g., \citealt{wanajo2014}). Detailed neutrino Boltzmann transport is currently not possible for compact binary merger simulations, and approximations must be made (e.g., \citealt{fujibayashi2017, kyutoku2018}). Furthermore, the neutrino luminosities and energies depend on the uncertain equation of state (e.g., \citealt{sekiguchi2015, 2017PhRvD..96l4005B}). Improvements in all those aspects is necessary in order to reliably constrain the range of electron fractions in all ejecta components of compact binary mergers. In addition, magneto-hydrodynamical effects can influence the behaviour of the dynamical and disk ejecta, but only a few codes currently include them \citep{palenzuela2015, ciolfi2017,2018ApJ...858...52S}. 

A further challenge is the need for high spatial resolution simulations, as all relevant features of the merger need to be resolved (e.g., Kelvin-Helmholtz instabilities at the contact interface). Simulations also need a high temporal resolution, as the ejecta leave the computational domain after a few milliseconds only. In the rare cases where the ejecta were followed for a longer time, the ejecta distribution at the end of the initial simulation was mapped to a different code \citep{rosswog2014, roberts2017, martin2018}, in order to study the impact of different heating sources on the dynamical ejecta.

Consistent long-term simulations, going up to about one second or longer, with an accretion disk and the emergence of a hypermassive neutron star that eventually collapses into a black hole  would considerably improve the predictive power of compact binary merger simulations. Further investigations of the impact of nuclear heating on the dynamical ejecta \citep{rosswog2014} as well as further increasing the parameter space of NS-NS and NS-BH merger simulations (masses, spins, etc.) are also promising paths to follow in the future.

\subsection{Simulations of Core Collapse Supernovae}
\label{sec:hydro_sim_sne}
The initial conditions of MR~SN simulations come from previously calculated stellar evolution models and these include several uncertainties and simplifications. Due to resulting uncertain magnetic field strengths, the magnetic field is artificial enhanced or decreased in MR~SN simulations. In addition, there are small scale turbulences and instabilities such as the magneto-rotational instability (e.g., \citealt{2009A&A...498..241O}) that can amplify the magnetic field. But to resolve such small scale effects, high-resolution and thus computationally-expensive simulations are needed (\citealt{moesta2017,nishimura2017}). Studies that follow explosions for one or two seconds with 3D simulations including detailed neutrino transports and general relativity will help to understand the dynamic of MR~SNe in greater details.

\subsection{Impact of Natal Kicks}
\label{sec:natal}
Something we have not included in our study is the natal kicks imparted onto neutron stars after the explosion of their progenitor stars (see \citealt{2017ApJ...846..170T} for a review). Depending on the coalescence timescales of NS-NS mergers and on the velocity of those kicks, it is possible for NS-NS binaries to merge outside galaxies (e.g., \citealt{1999MNRAS.305..763B,1999ApJ...526..152F,2006ApJ...648.1110B,2009ApJ...705L.186Z,2010ApJ...725L..91K,2014ApJ...792..123B,2017MNRAS.471.4488S}).  The simulations of \cite{2017MNRAS.471.2088S} showed that the spatial location of NS-NS mergers can play an important role on the amount of r-process material recycled into stars.

If the NS-NS mergers that occur outside galaxies are the ones that have the longest coalescence times, NS-NS mergers with short coalescence times would then be the ones that actively participate to the enrichment of Eu inside galaxies.  From the point of view of galactic chemical evolution modeling, this would be similar to using a DTD function truncated at some given timescale.  If that timescale is similar to or below the delay time needed for SNe~Ia to appear in the lifetime of stellar populations, this would be sufficient to reproduce the decreasing trend of [Eu/Fe] in the Galactic disk. In that case, however, the total number of NS-NS mergers would need to be increased in order to account for the ones that inject Eu outside the star-forming regions.  This scenario is to be confirmed or disproved by future studies.

\subsection{Actinides}
\label{sec:disc_actinides}
Actinide abundances in metal-poor stars could hold the key to disentangle the hydrodynamical environment(s) of the r-process. The existence of stars with super- and sub-solar actinide yields \citep{schatz2002, 2009ApJ...698.1963R, 2018arXiv180511925H, 2018ApJ...856..138J} possibly indicate varying r-process conditions. The fact that all currently known actinide-boost stars are very metal-poor ([Fe/H]~$\lesssim -2.0$) is another interesting finding. The fraction of actinide-boost stars and the Th/Eu ratios in metal-poor stars could potentially be used as an additional diagnostic tool to probe the contribution of different r-process sites.
Astrophysical models of dynamical and disk ejecta in neutron star mergers predict suitable environments for actinide production, and also several MR~SN models can provide the required conditions. Progress is needed before being able to quantify the typical mass of actinides ejected by NS-NS mergers and MR~SNe (see sections~\ref{sec:hydro_sim_cbm}~and~\ref{sec:hydro_sim_sne}). 

There might also be a possibility to infer actinide yields from direct kilonova observations. \cite{barnes2016} and \cite{rosswog2017b} showed that actinides undergoing $\alpha$-decay could represent an important contribution to the total nuclear heating powering the kilonova from a few days after the merger onwards.
Furthermore, \cite{2018arXiv180609724Z} showed that, given its half-life of $\sim$\,60\,days, the fission of $^{254}$Cf could be detectable several days after a NS-NS merger event, and could potentially be seen in the middle infrared band of future kilonova lightcurves.



\subsection{Possible Production of Europium by the i-Process}
\label{sec:disc_i_proc}
The r-process solar residual pattern does not take into account the potential contribution of the i-process \citep{cowan:77}.
The i-process can be activated at all metallicities in massive stars \citep[even down to metal-free stars, e.g.,][]{2018ApJ...865..120B,clarkson:18}, in rapidly accreting white dwarfs \citep[][]{denissenkov:17}, and in AGB and post-AGB stars of different types \citep[e.g.,][]{herwig:11,jones:16}. Observations of possible i-process abundance signatures have been detected in metal-poor stars \citep[e.g.,][]{Dardelet2014,lugaro:15,abate:16,Hampel2016,2016ApJ...821...37R,roederer:16} and in young open stellar clusters \citep[][]{mishenina:15}, for different elements  including Eu all the way up to Pb.

In the simulations of \cite{bertolli:13} and \cite{Dardelet2014}, considering an i-process activation between $\sim$~1 hour and $\sim$~1 day, the $^{151}$Eu/$^{153}$Eu isotopic ratio varies between 0.7 and 1.5, which is close to the solar value of 0.92. In spite of this interesting result, it does not necessarily mean that the i-process significantly contributed to the total amount of Eu found in the Milky Way. Because of the complex hydrodynamic simulations required to constrain the stellar conditions where the i-process takes place \citep[e.g.,][]{herwig:14,woodward:15}, and because of the large uncertainties in nuclear physics (\citealt{2018ApJ...854..105C,2018JPhG...45e5203D}), it is currently difficult to quantify the i-process contribution to the solar composition. In addition, large uncertainties in r-process nucleosynthesis calculations (e.g., \citealt{2016PrPNP..86...86M,2017ARNPS..67..253T,2018arXiv180504637H}) do not allow at the moment to derive strong constraints on the i-process component.

The conclusions presented in this paper are drawn under the assumption that the i-process does not contribute to the current inventory of Eu in our Galaxy. More investigations are needed to validate or invalidate this assumption.
\\
\section{Conclusions}
\label{sec:conclusions}
In this work, we compiled and analyzed a series of observational evidence that probe the properties of r-process sites, and addressed them using nucleosynthesis calculations, population synthesis models, and galactic chemical evolution simulations. At the moment, we cannot build a consistent picture when assuming that there is only one r-process site for elements between the second and third r-process peak. The list of agreements and inconsistencies are shown in Table~\ref{tab_summary}.

It is not possible to reproduce the decreasing trend of [Eu/Fe] in the Galactic disk with NS-NS mergers only using a DTD function in the form of $t^{-1}$ for the distribution of their coalescence timescales (see Sections~\ref{sect_common_message} and \ref{sec:scenario_2}).  This, however, is the form currently predicted by population synthesis models (see Section~\ref{sec:pop_synth}) and inferred from short GRB detections (see Section~\ref{sec:obs_GRB_DTD}).  We can reproduce the decreasing trend by steepening the slope of the DTD function (see also \citealt{2017ApJ...836..230C} and \citealt{2018arXiv180101141H}), or by adding a burst of NS-NS mergers before the onset of SNe~Ia on top of a $t^{-1}$ distribution (see Section~\ref{sec:scenario_3}). But this is not consistent with the similar distribution of short GRBs and SNe~Ia in early-type galaxies (see Section~\ref{sec:obs_GRB_host}). Assuming a constant short delay time for NS-NS mergers is inconsistent with several observations (see Section~\ref{sec:scenario_1}), including the detection of GW170817 in an early-type galaxy. 

Because nucleosynthesis calculations show that NS-NS mergers are able to produce a robust r-process pattern (see Appendix~\ref{sec_nucleo_nsm}), and because the merger rate established by LIGO/Virgo is significantly high (see Section~\ref{sec:obs_grav_rate}), NS-NS mergers are likely to play an important role in the evolution of r-process elements. If that is the case, and if we assume current observational constraints are all reliable, then one solution is to involve a second r-process site in the early universe capable of producing Eu (see Section~\ref{sec:scenario_5}). This extra site should fade away at later times, at higher metallicity (see pink dashed line in Figure~\ref{fig:knee}), and would roughly account for $\sim$\,50\,\% of the Eu produced in the early universe, and $\sim$\,10\,\% of all Eu currently present in the Milky Way. This is a plausible solution, but it does not mean it is the correct solution.

MR~SNe could be this extra site of Eu. To reach the third r-process peak and produce Eu along the way, these SNe need very strong magnetic fields (see Appendix~\ref{sec:cc_sne_robust}). Because massive stars are likely to rotate faster at low metallicity, the magnetic field present during a MR~SN could also be higher at low metallicity, which would explain why this second site fades away at higher metallicity (see Section~\ref{sec:scenario_5}). This conclusion is based on the chemical evolution of the Galactic disk, which encodes most of the evolution history of the Milky Way. We note that any other production site of Eu that respect this metallicity dependency is a possible candidate.

Although we presented a possible solution to the current inter-disciplinary tensions, the information shown in Table~\ref{tab_summary} should be taken as the starting point for future work.  Other possible solutions exist (e.g., Sections~\ref{sec:natal} and \ref{sec:disc_i_proc}), and further observations might change our interpretation of the observational evidence listed in Section~\ref{sec:obs}.  In particular, more detections of NS-NS mergers by LIGO/Virgo and improvements in the interpretation of kilonova lightcurves will better define how much room there is for other r-process sites.

\acknowledgments
We are thankful to Erika Holmbeck, and Ian Roederer for discussions on the abundances of metal-poor stars, to Gabriele Cescutti for discussions on modeling neutron-capture elements in a chemical evolution framework, and to Mohammadtaher Safarzadeh for discussions on the impact of natal kicks. This research is supported by the ERC Consolidator Grant (Hungary) funding scheme (project RADIOSTAR, G.A. n. 724560) and by the National Science Foundation (USA) under grant No. PHY-1430152 (JINA Center for the Evolution of the Elements). This paper included work from the ChETEC (Chemical Elements as Tracers of the Evolution of the Cosmos) COST Action (CA16117), supported by COST (European Cooperation in Science and Technology), and has benefited from discussions at the 2018 Frontiers in Nuclear Astrophysics Conference supported by the JINA Center for the Evolution of the Elements. AA, ME, and MR were supported by the ERC starting grant “EUROPIUM” (Grant No. 677912) and BMBF under grant No.05P15RDFN1. ME is also supported by the Swiss National Foundation (Project No.~P2BSP2\_172068). AF is partially supported by NSF-CAREER grant AST-1255160 and NSF grant 1716251. MP acknowledges the support of STFC through the University of Hull Consolidated Grant ST/R000840/1 and ongoing resource allocations on the University of Hulls High Performance Computing Facility viper. KB acknowledges support from the Polish National Science Center (NCN) grant: Sonata Bis 2 DEC-2012/07/E/ST9/01360. FM was supported by funds from Trieste University (FRA2016).

\software{\texttt{JINABase} \citep{2017arXiv171104410A},
          \texttt{OMEGA} \citep{2017ApJ...835..128C},
          \texttt{SYGMA} \citep{2017arXiv171109172R},
          \texttt{StarTrack} \citep{Belczynski2002,Belczynski2008},
          \texttt{matplotlib} (\url{https://matplotlib.org}),
          \texttt{NumPy} \citep{2011arXiv1102.1523V}.
          }

%

\vspace{5mm}

\appendix

\section{Minimum Delay Time and Star Formation History}
\label{app:gce}
\begin{figure}
\center
\includegraphics[width=3.35in]{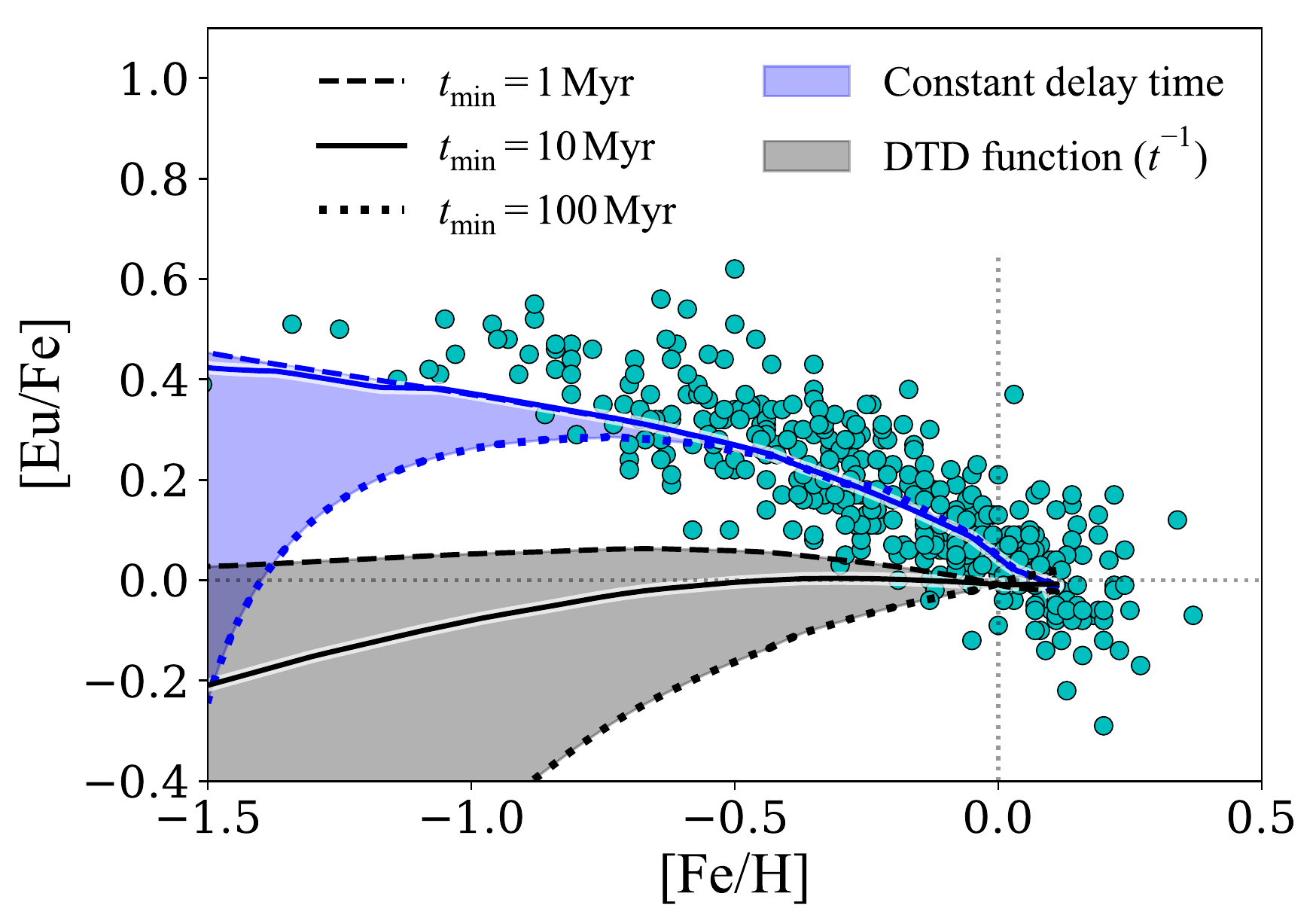}
\caption{Impact of using different minimum delay times for NS-NS mergers on the predicted chemical evolution of Eu. The blue and black bands show the prediction when assuming constant delay times or delay-time distribution (DTD) functions, respectively. The dashed, solid, and dotted lines show the predictions using different minimum delay times between the formation of the progenitor stars and the onset of NS-NS mergers. In the case of constant delay times, the minimum delay times are the constant delay times. Cyan dots are stellar abundances data for disk stars taken from \cite{2016A&A...586A..49B}. The dotted black lines mark the solar values (\citealt{2009ARA&A..47..481A}).} \label{fig:app_t_min}
\end{figure}

In Section~\ref{sect_common_message}, we mentioned that GCE predictions for [Eu/Fe] at [Fe/H]\,$>$\,$-1$ are insensitive to the minimum delay time of NS-NS mergers, when adopting the constant delay-time approximation. In this section, we test this statement when adopting a DTD function in the form of $t^{-1}$, using $t_\mathrm{min}=1$, 10, and 100\,Myr, where $t_\mathrm{min}$ refers to the minimum delay time between the formation of the progenitor stars and the merger event. We note that according to the population synthesis predictions presented in Figure~\ref{fig:dtd_nsns}, $t_\mathrm{min}$ is typically below 100\,Myr. As shown in Figure~\ref{fig:app_t_min}, using the GCE code \texttt{OMEGA}, the predicted trends at [Fe/H]\,$>$\,$-1$ are similar when using a constant delay time (blue band, see also \citealt{2014MNRAS.438.2177M,2015A&A...577A.139C,2017ApJ...836..230C}), but are significantly different when using a DTD function (black band). Nevertheless, varying $t_\mathrm{min}$ can still not provide a good agreement with the measured [Eu/Fe] trend when using a DTD function in the form of $t^{-1}$. This parameter therefore do not affect the general conclusions of this work, although it could affect the estimated contribution of the hypothetical extra r-process site (see Section~\ref{sec:scenario_5}).

\begin{figure}
\center
\includegraphics[width=3.35in]{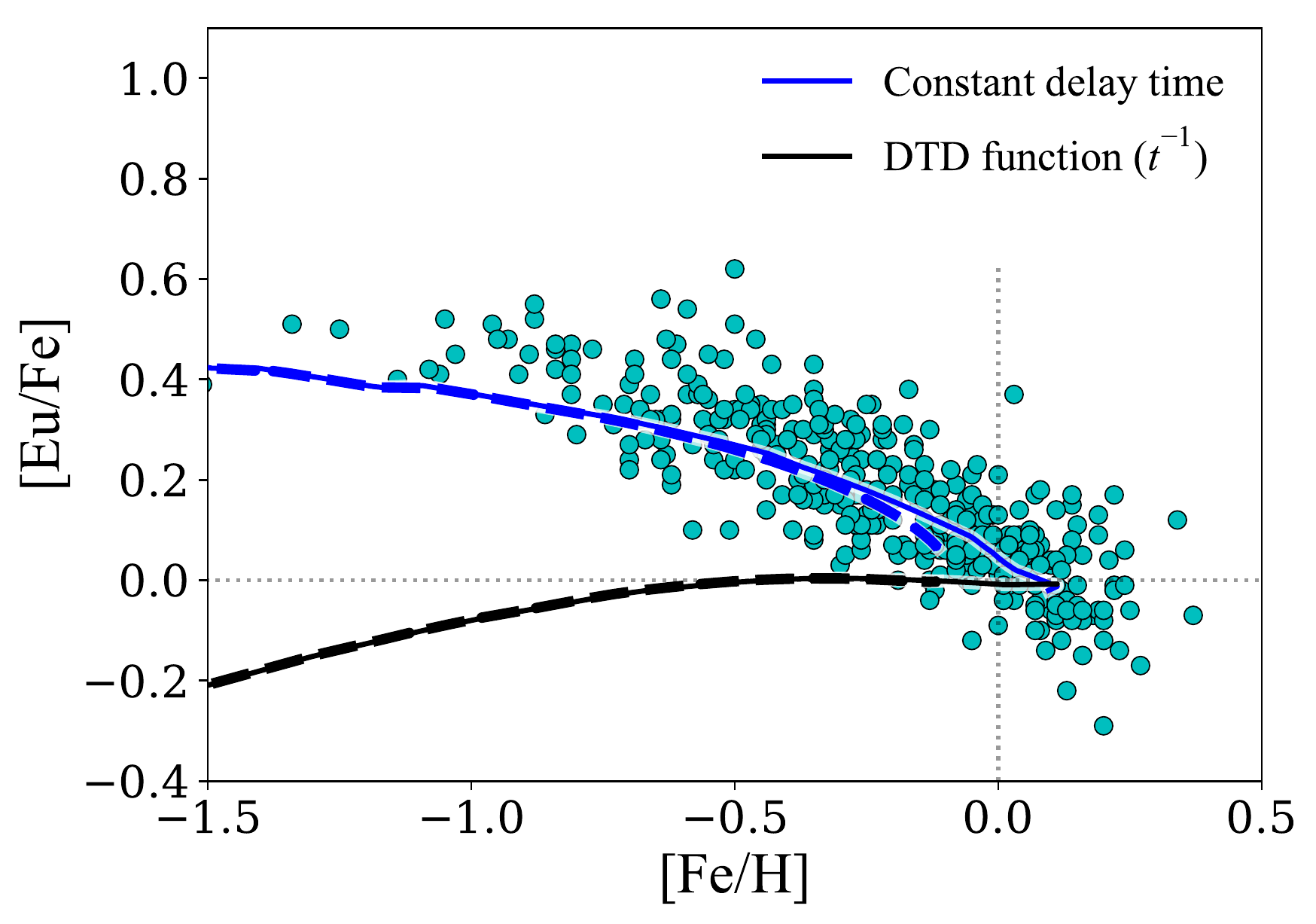}
\caption{Impact of using different star formation histories on the predicted chemical evolution of Eu with NS-NS mergers. The blue lines show predictions when assuming that NS-NS mergers all occur after a constant delay time of 10\,Myr following the formation of progenitor stars, while the black lines shows predictions when assuming a delay-time distribution (DTD) function in the form of $t^{-1}$. The solid lines show results with a decreasing star formation history, from 11\,M$_\odot$\,yr$^{-1}$ at time zero down to 2\,M$_\odot$\,yr$^{-1}$ after 13\,Gyr, while the dashed lines show results with a constant star formation history of 6\,M$_\odot$\,yr$^{-1}$. All simulations generate a current stellar mass of 5$\times$10$^{10}$\,M$_\odot$. Cyan dots are stellar abundances data for disk stars taken from \cite{2016A&A...586A..49B}. The dotted black lines mark the solar values (\citealt{2009ARA&A..47..481A}).} \label{fig:app_sfh}
\end{figure}

As a complement, we also explored the impact of the star formation history (see also \citealt{2018arXiv180101141H}). Figure~\ref{fig:app_sfh} shows the predictions made by the \texttt{OMEGA} code when using a decreasing star formation history (solid lines) and a constant star formation history (dashed lines). Overall, we find that the effect of the star formation history is minor in comparison to the effect of using different delay time assumptions for NS-NS mergers. As in Figure~7 of \cite{2018arXiv180101141H}, the maximum [Fe/H] reached by our predictions is shifted to lower [Fe/H] values when using a constant star formation history (dashed lines), likely because of the larger amount of infalling gas compared to the decreasing star formation history, which has a current star formation rate three times lower than in the constant case. Overall, as for the minimum delay time for NS-NS mergers, the adopted star formation history should not affect the general conclusions of this work.

\section{Population Synthesis: Brief Description of Adopted Models}
\label{app:popsynth}

Here we discuss the properties of the population synthesis models adopted for our study in more detail (see section~\ref{sec:pop_synth}).
We use the upgraded population synthesis code {\tt StarTrack} \citep{Belczynski2002,Belczynski2008}. Improvements relevant for massive stars include a better treatment of the common envelope evolution \citep{2012ApJ...759...52D} and a remnant mass prescription \citep{Fryer2012} that reproduces the first gap found in between the observed mass distributions of neutron stars and black holes in the Milky Way \citep{Belczynski2012}.  Effects of pair-instability pulsation SNe and pair-instability SNe are taken into account, as they are believed to create the second mass gap in the observed mass distribution of compact objects \citep{Belczynski2016c}.

We account for the formation of neutron stars via electron-capture SNe (e.g., \citealt{Miyaji1980,Podsiadlowski2004,2018MNRAS.474.2937C}) and through accretion-induced collapse of white dwarfs \citep[e.g.,][]{Nomoto1991,Ruiter2018}.  The maximum mass of neutron stars is set to $2.5$\,M$_\odot$, which is consistent with the heaviest known neutron stars of $2.3$\,M$_\odot$ (\citealt{Linares2018}). The initial conditions of the binary systems are constrained by observations \citep{deMink2015}.
The differences between the models can also be seen in Figures~\ref{fig:dtd_nsns}~\&~\ref{fig:dtd_bhns} in the main body of this article.
Model M10 represents the standard input physics in {\tt StarTrack}.  In particular, the natal kick velocities initially imparted on black holes are small, and in some cases non-existent. For neutron stars, the natal kicks are defined by a Maxwellian distribution with $\sigma=265$\,km\,s$^{-1}$.  However, the exact values of the natal kicks can be altered by the amount of material falling back onto the compact objects after the SN explosions (see \citealt{Belczynski2018b} for details).  The magnitude of natal kicks is an important quantity, as it can significantly modify the coalescence timescales of NS-NS and BH-NS systems (e.g., \citealt{1999A&A...346...91B}).

The M13 model is similar to the M10 model, but the Maxwellian natal kick distribution with $\sigma=265$\,km\,s$^{-1}$ is adopted for both black holes and neutron stars. In addition, the amount of fallback does not regulate the kick velocities. We note that neutron stars formed by electron-capture SNe and by the accretion-induced collapse of white dwarfs still have natal kicks set to zero, as it is the case for all models presented in this section.

In model M20, the natal kick prescription for black holes and neutron stars is the same as in the M10 model, but the input physics describing the evolution of binary systems during the common envelope phase is modified.  This includes for example 80\,\% non-conservative Roche Lobe overflow and 5\,\% Bondi-Hoyle rate accretion, as opposed to 50\,\% and 10\,\%, respectively, in the M10 model (see \citealt{Belczynski2018b} for details).

The M23, M25, and M26 models have the same modified input physics as in the M20 model, but adopt a Maxwellian natal kick velocity distribution for both neutron stars and black holes, as in Model M13.  The dispersion velocity $\sigma$ for models M23, M25, and M26 is 265, 130, and 70\,km\,s$^{-1}$, respectively.

\section{R-Process Nucleosynthesis in Different Environments}
\label{sec_nucleo}

\subsection{Neutron Star Mergers}
\label{sec_nucleo_nsm}

In NS-NS mergers, there are various ways of ejecting matter and those lead to different conditions and nucleosynthesis. The total amount of ejecta depend on the configuration of the binary system (e.g., neutron star masses, eccentricity of the orbit). For the dynamical ejecta, between about $10^{-3}$ and $10^{-2}$~M$_{\odot}$ of very neutron-rich (low $Y_e$) material can be ejected by means of tidal effects or shocks in the interaction zone between the two neutron stars. These numbers are in agreement with estimates of the ejected mass for GW170817 (see Section~\ref{sec:obs_grav_kilo}), and are large enough to recover the scatter seen in the [Eu/Fe] abundances of metal-poor stars (see Section~\ref{sec:obs_scatter}).

The dynamical ejecta robustly produce the heavy r-process nuclei from the second peak up to the actinide region with mass numbers A above 200. The extreme neutron-richness of the dynamical ejecta leads to fission cycling and an r-process path close to the neutron drip line (further favoured by rapid expansion/cooling). As a consequence, such environment only produces nuclei from the typical fission fragment mass numbers $A \approx 120$ and larger. The do not produce the lighter first-peak r-process nuclei. NS-NS mergers are thus consistent with the robust main r-process patterns seen in metal-poor stars (\citealt{2008ARA&A..46..241S}, see also Section~\ref{sec:obs_met_poor}). 

However, other hydrodynamical codes lead to broader ranges in electron fraction and entropy \citep{bauswein2013,sekiguchi2015,foucart2016,radice2016,2017PhRvD..96l4005B}. Simulations incorporating general relativity effects exhibit a more energetic collision, since the neutron star radii are smaller. This leads to strong shocks that increases the $Y_e$ and the entropy in the ejecta. But even these simulations produce robust abundances between the second and third r-process peak, since the ejecta consist of two major components: those that stop at or before the second peak, and those that produce the full mass range of r-process nuclei up to the actinides in a robust way. Some recent works have found conditions with low $Y_e$ and large entropies (related to shocked ejecta with very fast expansion velocities) which result in an unusual abundance pattern (reported, e.g., in \citealt{metzger2015,barnes2016,2017PhRvD..96l4005B}). However, these conditions contribute to only a very small fraction of the total ejecta, and therefore have no real effect on the integrated abundances.

In addition to the dynamical ejecta, there are other mass loss channels operating on longer timescales that are related to the formation of a post-merger disk around the central remnant object \citep{beloborodov2008,dessart2009,foucart2012, fernandez2013b, perego2014,fernandez2015,just2015,martin2015,siegel2017}. Again, the formation of the disk and its mass depend on the binary system configuration. If the central object does not promptly collapse to a black hole, it emits neutrinos that, together with the neutrinos emitted from the disk, can generate a mass outflow similar to neutrino-driven winds in CC~SNe \citep{perego2014,martin2015, just2015}. Simulations also show that disk material can become gravitationally unbound by means of viscous effects and nuclear recombination \citep{fernandez2013b,just2015,fernandez2015}. Due to the longer timescales of these ejecta, they are exposed to neutrinos for a longer time compared to the dynamical ejecta. And because neutrinos transform neutrons into protons, these longer-timescale ejecta are typically less neutron-rich.

While the viscous ejecta can still produce the heaviest r-process nuclei \citep{wu2016}, the neutrino-driven ejecta have been found to produce first-peak elements only \citep{perego2014,martin2015}. The viscous and the neutrino-driven ejecta represent about 5-20\,\% \citep{fernandez2013b,just2015,fernandez2015} and 5\,\% \citep{martin2015} of the initial disk mass, respectively. These numbers depend on the fate of the central remnant. For example, a quicker collapse into a black hole leads to a smaller ejected mass via these two mass loss channels. Finally, magnetic fields can also contribute to the acceleration and ejection of mass (\citealt{siegel2017}). However, only preliminary results are available and it is still early to make conclusive statements about the importance of magnetic fields.

\subsection{Core-Collapse Supernovae}
\label{sec:cc_sne_robust}

At the surface of a proto-neutron star, electron capture can deleptonize the material.  If this material is ejected, it can produce r-process elements.
But in standard CC~SNe, the matter is mainly ejected by neutrinos, which reduce the number of neutrons by converting them into protons. 
In such conditions, current simulations indicate that standard CC~SNe could at most produce trans-iron elements like Sr, Y, Zr, Mo, and Ru (first-peak elements), and perhaps in some cases elements up to Ag (e.g., \citealt{roberts:10,Arcones2011,Eichler2018,Bliss2018} and references therein). In order to produce heavier elements such as Eu, another ejection mechanism is necessary. 
Magnetic fields and rotation have been identified as a possibility by \cite{cameron2003} and \cite{nishimura2006}. Simulations of magneto-rotational (MR) SNe have become possible at least with simplifications in the neutrino transport. We refer to the introduction of Section~\ref{sec:nucleo} for a list of other possible r-process sites.

Based on three-dimensional simulations with a simple leakage treatment for the neutrinos, \cite{winteler2012} found that the heavy r-process can be synthesized in MR~SNe. The two-dimensional simulations of \cite{nishimura2015,nishimura2017} with parametric neutrino treatment found that when the magnetic field is strong, and neutrino emission is reduced, a jet-like explosion develops and neutron-rich material is ejected. Again with a strong magnetic field, \cite{moesta2017} have shown, however, that in 3D simulations it becomes harder to reach the neutron-richness required for a robust r-process. They also confirm the results of \cite{nishimura2015}, demonstrating that when the neutrino emission is artificially increased, one can go from producing a robust r-process to producing a weak r-process up to the second peak only.

This behavior could potentially explain the abundance patterns of metal-poor stars, such as HD~122563 (\citealt{2006ApJ...643.1180H}), enhanced in Sr and other elements at the neutron shell closure $N=50$ relative to Ba. Furthermore, because MR~SNe might synthesize a variety of r-process abundance patterns depending on the conditions, from a week to a robust r-process including the second and third peaks, such SNe could explain the variety of metal-poor stars having abundance patterns in between stars like HD~122563 and stars showing the full r-process pattern (see \citealt{2010ApJ...724..975R}). Improved neutrino transport is mandatory before making any definite conclusions, since neutrinos can lead to a broad variability in the conditions of the matter ejected \citep{obergaulinger2017,reichert_inprep}. 

Regarding the total amount of r-process material ejected, MR~SNe may contribute very differently depending on the strength of the magnetic field. In general, the ejected mass containing r-process elements will be larger than with neutrino-driven winds in regular CC~SNe, but will be smaller than with NS-NS mergers. This kind of explosions is expected to occur in a few percent of all SNe at most, preferentially at low metallicities where the magnetic field could be amplified by faster stellar rotation velocities. Because of this possible dependence on metallicity, MR~SNe represent an interesting candidate for our multiple r-process sites scenario presented in Section~\ref{sec:scenario_5}.


\bibliographystyle{yahapj}
\bibliography{apj-jour,ms}


\end{document}